\documentclass[%
 aip,
 amsmath,amssymb,
reprint,%
]{revtex4-2}

\usepackage{graphicx}% Include figure files
\usepackage{dcolumn}% Align table columns on decimal point
\usepackage{bm}% bold math
\usepackage[utf8]{inputenc}
\usepackage[T1]{fontenc}
\usepackage{mathptmx}
\usepackage{etoolbox}
\usepackage{xcolor}
\usepackage{accents}
\usepackage{soul}
\usepackage{graphicx}% Include figure files
\usepackage{dcolumn}% Align table columns on decimal point
\usepackage{bm}% bold math
\usepackage{xcolor}
\usepackage{floatrow}
\usepackage{graphicx}
\usepackage{subfig}
\usepackage{caption}
\usepackage{setspace}
{\color{red}}%
{}
\makeatletter
\def\@email#1#2{%
 \endgroup
 \patchcmd{\titleblock@produce}
  {\frontmatter@RRAPformat}
  {\frontmatter@RRAPformat{\produce@RRAP{*#1\href{mailto:#2}{#2}}}\frontmatter@RRAPformat}
  {}{}
}%
\makeatother
\begin{document}

\preprint{To appear in Journal of Rheology: \url{https://doi.org/10.1122/8.0001241}}

\title[]{Multi-scale Modeling of the Electro-viscoelasticity of Charged Polymers in Combined Flow and Electric Fields}
% New title suggestion: A Multiscale Viscoelastic Model of Charged Polymers in Flow with Constant Electric Fields
%\title{Multi-scale Modeling of the Viscoelastic Response of Charged Polymers in Combined Flow and Electric Fields}
\author{Zachary Wolfgram}

\affiliation{ 
Department of Mechanical Science and Engineering, University of Illinois at Urbana-Champaign, Urbana, IL 61801, USA
}%
\author{Jeffrey G. Ethier}%
 \email{jeffrey.ethier.1@afrl.af.mil}
 \thanks{Co-corresponding authors}
\author{Matthew Grasinger}
 \email{matthew.grasinger.1@afrl.af.mil}
 \thanks{Co-corresponding authors}
\affiliation{%
Materials \& Manufacturing Directorate, Air Force Research Laboratory, Wright-Patterson AFB, OH, 45433, USA
}%

\begin{abstract}
The behavior of polymers in combined flow and electric fields underlies many manufacturing processes but remains poorly understood. We model charged polymers across scales, extending the Rouse model for a bead-spring chain to include a charge density distributed along the polymer chain, and derive the viscoelastic stress under homogeneous shear and electric fields. The viscosity increase depends on field-flow orientation, scaling quadratically with select components of the electric field strength, modulated by the effective charge sequence relaxation time and dielectric constant. Inspired by this, a new continuum model--the upper-convected electro-Maxwell (UCEM) model--is proposed. It resembles an upper-convected Maxwell model with polarization stresses expressed through an electric field dyadic subject to upper-convected time derivatives. The UCEM model captures the electro-rheological coupling for unentangled, salt-free melts at low shear rates and remains predictive into the moderately nonlinear regime ($Wi \approx 2-4$) provided its parameters are treated as state-dependent material functions. We analyze the constitutive response for various flows under applied electric fields, discussing model limitations and demonstrating compliance with the second law of thermodynamics. Coarse-grained molecular dynamics (MD) simulations of Kremer-Grest chains confirm distinct relaxation timescales for chain dynamics versus charge redistribution, consistent with UCEM predictions. Critically, the upper-convected time derivative of the electric field dyadic is required in the stress evolution to account for stretching and rotation of charge pairs in flow. This reproduces the viscosity scaling observed in both the Rouse and MD results, whereas standard continuum formulations without these terms fail to capture this scaling.
\linebreak
\\
(\textit{To appear in Journal of Rheology:} \url{https://doi.org/10.1122/8.0001241})
\end{abstract}

\maketitle

\section{Introduction}

Electromagnetic fields are a key input in various manufacturing applications for soft and multifunctional materials, composites, and fluids. For instance, electrowetting manipulates the surface tension of a dielectric liquid via an applied electric field, causing electric dipoles to align and reduce surface tension orthogonal to the field direction. This enhances wetting and has wide-ranging applications in microfluids~\cite{li2020current}. In the context of composites processing, electrowetting promotes infiltration of polymer resins into the sub-micron crevices of materials like woven carbon fiber reinforcement, mitigating defects and enhancing mechanical strength~\cite{chen_enhancing_2022}. Increasing the electric field intensity can drive electrospinning~\cite{larrondo_electrostatic_1981,winslow_induced_1949,kakade2007electric}, which produces long, thin polymeric fibers (often on the $\mu m$ or $n m$ scale) with high strength and porosity. Optimizing these fields can align the polymers during crystallization, thereby increasing the shear modulus and electrical conductivity~\cite{kakade2007electric}. Further applications include (1) electrospraying~\cite{green2024self,park2023efficient} for achieving even coatings, and (2) using electromagnetic fields as a powerful tool for programming the structure and alignment of electromagnetically responsive fillers~\cite{arguin2015electric}, leading to enhancements in electrical conductivity and mechanical properties in composites.

Understanding the viscoelasticity and flow of polymers in electromagnetic fields is essential because many manufacturing processes for polymer-based materials and technologies (e.g. soft actuators, soft robotics, and energy harvesters~\cite{BEHERA2021104369,grasinger2021nonlinear}) are often electrically-driven or electrically-assisted. Electro-viscoelasticity is an extension of electrorheology (ER) that focuses on the impact electric fields have on the deformation properties of a medium possessing both fluid and solid-like characteristics~\cite{BEHERA2021104369,MEHNERT2021104625,giorgi_modelling_2023}. The electrorheological effect describes the phenomenon where electric fields interact with fluid flow, potentially breaking isotropy and inducing additional internal stresses and torques~\cite{grasinger2021nonlinear,ruzicka_electrorheological_2000,prusa_flow_2012,ruzicka_modeling_2004}. ER technologies have been shown to reversibly tune viscosity and other material characteristics in ER fluids and elastomers\cite{liang2023efficient}.
Furthermore, recent work on charge block polyampholytes \cite{shi_influence_2023} brings into question the underlying physics of these polymers and flow response behavior.  

Extensive research has been dedicated to understanding the rheology of charge-containing polymers, primarily focusing on ionomer melts and polyelectrolyte coacervates. In ionomer melts, sparse ionic groups aggregate into physical crosslinks or clusters that increase relaxation times, alter linear viscoelastic behavior, and drive extensional strain hardening.\cite{vanhoorne1996low,ling2012linear,greener1993melt,weiss2009rheological} Conversely, in solution, the rheology of polyampholytes and polyelectrolyte complexes is governed by the continuous breaking and reforming of temporary intra- and inter-chain electrostatic interactions. This leads to complex viscoelastic behaviors such as coacervation and physical gelation where the viscosity and dynamic moduli are highly tunable by salt concentration, charge ratio, and solvent quality\cite{bossard2004rheological,hamad2018linear,li2022effect,zhang2022combined,peiffer1986synthesis,zhang2005thickening} However, despite these advances, the rheological behavior of weak polyampholyte-like melts where dense, alternating charge sequences interact within a solvent-free, highly entangled or unentangled environment, remains largely unexplored. Understanding the melt rheology of these sequence-defined polyampholytes is particularly critical when they are subjected to combined macroscopic flow and external electric fields, as the field-driven orientation of the charges along the backbone introduces new relaxation timescales that are not captured by traditional ionomer or polyelectrolyte models.

The viscoelastic behavior of polymers has been studied for decades, as it is fundamental to the processability and characteristic behavior of the material. Theoretical and multi-chain level models, often grounded in statistical mechanics, provide a framework for predicting viscoelastic behavior from molecular-scale features.\cite{likhtman20121} For instance, seminal work by Rouse idealized non-interacting polymer chains as bead-spring systems, expanding on the Hookean dumbbell model. The linear viscoelastic stress was then shown to be a sum of the stress contributions from individual internal relaxation (Rouse) modes.~\cite{rouse1953theory,doi1996introduction}.
This early work was later extended to incorporate hydrodynamic interactions~\cite{zimm1956dynamics}, among other physics not captured in the Rouse model.
The dielectric chain description given by Zimm \cite{zimm1956dynamics} was also expanded within Adachi \cite{adachi_dielectric_1993}, with a focus on broader polarization responses of dipole moments of the beads themselves, and dipole moments between bead charges.
Martin and Anderson derived a viscoelastic model focusing on the alignment of a flexible chain with a field-induced bead dipole moment \cite{10.1063/1.471176}.
Our work here can be considered an extension of the original Zimm approach, focusing on dipole moments between bead charges, in that we derive a chain-level description of a polymer with an oscillating charge sequence (e.g., a polyampholyte).
While previous models consider the competition between a single electrostatic energy and entropy, we explore how polarizable segments along the chain contribute to the viscoelastic stress.
While initially derived for polymer chains in dilute solution, the Rouse model has been shown to capture the linear viscoelastic behavior of unentangled polymer melts. 
For large molecular weights above the entanglement limit, the viscoelastic stress can no longer be described by the Rouse model. In such cases, other constitutive equations should be used such as the Doi-Edwards tube model and extensions thereof,\cite{edwards1988tube,mcleish2002tube,likhtman2003simple,likhtman2002quantitative} or discrete chain models such as the slip-link\cite{schieber2014entangled} and slip-spring models\cite{uneyama2012multi}. 

At the macroscopic scale, the linear viscoelastic behavior of polymers can be described, to first-order, by the upper-convected Maxwell (UCM) model, which predicts constant viscosity and first normal stress difference at low shear rates. However, one limitation of this model is its inability to capture the shear thinning behavior. Extensions of this model have been derived, such as the Oldroyd-B~\cite{oldroyd1950formulation} and Giesekus\cite{giesekus1982simple} models, that incorporate solvent (viscous) contributions or second-order terms, respectively. Motivated by these earlier works, we choose to extend the UCM model here, developing a minimal electrorheological constitutive model that is objective and reproduces the scaling features and orientational dependence observed in lower-scale models.

In addition to continuum-level models, coarse-grained molecular dynamics (MD) simulations present an alternative method to probe equilibrium and nonequilibrium rheology of polymer systems at the mesoscale, observing phenomena such as shear thinning and entanglement effects~\cite{Kremer_1990,likhtman2007linear}. Furthermore, these models have been recently used to study the dynamics and rheology of polyelectrolyte and polyampholyte-polyelectrolyte complexes.\cite{liang2022coarse,jeon2005molecular}
While past work has uncovered fundamental structure-rheology and structure-electroelasticity~\cite{grasinger2021nonlinear} relationships in polymers, few have considered how the evolution of polymer stress is affected by applied electric fields in charged polymer systems, producing different dynamics and viscoelastic behavior.

At the continuum scale, prior models developed for electrorheological fluids often coupled the shear and electric fields using the strain rate tensor and the electric field dyadic. For example, the Bingham model for yield-stress fluids\footnote{Past use of the Bingham plastic model is motivated by electrorheological suspensions, where particles chain under an applied electric field and exhibit a field-dependent yield stress. Such behavior is not generally observed in polymer solutions.} and other continuum models developed by Rajagopal, Yalamanchili, and Wineman are some of the earliest known models describing electrorheological behavior.\cite{bingham1922fluidity,rajagopal1996modeling} These known constitutive equations often lack the objectivity (i.e., frame indifference) required for a correct measure of fluid stress and are not directly connected to the structure and behavior of the underlying polymers, which limits insight for future material design~\cite{ruzicka_modeling_2004}. For example, a previous study showed that the direction of shear flow relative to the direction of the electric field modulates the electrorheological effect in ways that are not well explained by existing continuum theories.\cite{ruzicka_modeling_2004} 
In addition, experimental measurements~\cite{huo_electric_2020,abu-jdayil_effects_1996} report viscosities that current continuum scale theories fail to correctly characterize. There are many possible intracacies to the dynamics of polymer flow in an electromagnetic field. In dielectric fluids, the alignment of electric dipoles with the field competes against the rotation induced by fluid velocity gradients, a competition that results in an electric field-dependent viscosity~\cite{edamura2004electrorheology,pennington1969couette}.
%It suggested that the resulting extrema directions of viscosity arise from the competition between the excitation of a Rouse mode due to the shear rate and the excitation of another mode via the electric field. 

Here, we aim to develop a minimal, first-approximation multiscale framework that explains the distinct relaxation timescales observed for polyampholyte-like chains under combined flow and electric fields. We begin with a chain-level description of a charged polymer examined previously in the literature \cite{zimm1956dynamics,adachi_dielectric_1993}, modifying the overdamped Langevin equations for the Rouse model to incorporate a defined charge distribution along the polymer backbone that interacts with electrostatic forces from an externally applied electric field. Inspired from the derived viscoelastic stress in this model, a new continuum model is proposed: the upper-convected electro-Maxwell model, which resembles a modified UCM model and includes the evolution of polarization stresses in terms of an upper-convected time derivative of the electric field dyadic. The continuum model is developed for unentangled polymer melts at low shear rates, under the key assumption that long-range electrostatic interactions between charges are negligible. This assumption holds in the limit where charge-charge interactions are weak compared to charge-field interactions. Results from this idealized model are then compared to coarse-grained molecular dynamics (MD) simulations of bead-spring polymer chains in shear flow, which explicitly include these long-range Coulomb interactions. The MD results verify that, by including the upper-convected time derivative of the electric field dyadic, the viscosity scaling -- based on the difference between the overall relaxation and charge sequence relaxation time -- is correctly characterized. In addition, we show that, in contrast to standard electrorheological continuum models, the UCEM model satisfies the second law of thermodynamics, frame indifference, and recovers the directional dependence of the electric field as predicted by more detailed models. %Finally, a brief examination of dynamic electrorheological properties shows agreement with recent experimental data on a Poly(methyl methacrylate) (PMMA) melt~\cite{huo_electric_2020}.

\section{Molecular Dynamics Simulation Details}

We compare our constitutive model to the bulk stress response from molecular dynamics (MD) simulations of an unentangled polymer melt using the Large-scale Atomic/Molecular Massively Parallel Simulator (LAMMPS)\cite{Thompson_2022}.
The polymer chains are modeled using a standard Kremer-Grest bead-spring model, which represents each chain as a collection of monomers (beads) connected by springs, where non-bonded monomers are purely repulsive with each other \cite{Kremer_1990,Smook_2024,Aoyagi_2000,Mees_2023}.
To capture the non-linearity of the stiffness of the polymer under finite deformations, we apply the finite-extensible nonlinear elastic (FENE) potential for all bonded monomers defined by,
\begin{equation}
    U_{\mathrm{FENE}}\left(r\right) = -0.5KR_0^2\ln\left[1-\left(\frac{r}{R_0}\right)^2\right],
\end{equation}
where $K$ is analogous to the spring constant, $R_0$ is the maximum extensibility of the bonds, and $r$ is the current distance between the two monomers \cite{Kremer_1990}. 
All bonded and non-bonded monomer interactions are described by the Lennard-Jones (LJ) potential,
\begin{equation}
    U_{\mathrm{LJ}}\left(r\right) = \begin{cases}
        4\epsilon\left[\left(\frac{\sigma}{r}\right)^{12}-\left(\frac{\sigma}{r}\right)^6\right] & r<r_c, \\
        0 & \mathrm{otherwise} \\
    \end{cases}
    \label{LJ}
\end{equation}
where $\sigma$ and $\epsilon$ are the monomer size
and interaction energy well depth of the LJ potential, respectively, and $r_c$ is the cutoff distance. 
All monomers are a size of 1$\sigma$ and we set the cutoff distance to $r_c=2.5\sigma$ to include a small attraction between non-bonded monomers. 
Finally, we set $K=30\epsilon/\sigma^2$ and $R_0 = 1.5\sigma$ to prevent chains from crossing through each other. \cite{Kremer_1990}

To incorporate electrostatics, a pairwise Coulomb interaction potential is added of the form
\begin{equation}
    U_{C}\left(r\right)=\frac{Cq_iq_j}{\epsilon_{b}r} \quad \text{ if } r<r_c, \qquad =0 \quad \text{ otherwise}.
\end{equation}
where $q_i$ and $q_j$ are the charges of the interacting monomers, $C$ is the energy-conversion constant equal to $1/4\pi\epsilon_0$, and $\epsilon_{b}$ is the background dielectric constant set to 1\cite{Smook_2024,Thompson_2022}. It is important to note that $\epsilon_b$ is distinct from the total macroscopic dielectric permittivity ($\epsilon_r$) of the polymer melt. The total permittivity, which arises from the spatial fluctuations of the polymer charge sequences, dictates the continuum polarization susceptibility ($\epsilon_{dielec}$) that will be introduced in Sections III and IV.
We include the long-range Coulomb forces between polymer chains by allowing charge-charge interactions to be explicitly solved within $r_c\leq4.0\sigma$ and use the particle-particle particle-mesh (PPPM) method for calculating long-range forces with an error set to $10^{-4}$.
The force of the electric field on each charge is given by
\begin{equation}
    F^E_{i}=q_i E,
\end{equation}
where $E \in \mathbb{R}^3$ denotes the externally applied, homogeneous, electric field. Finally, the net force on each monomer $i$ is
\begin{equation}
    F_{i} = -\nabla_i \left[\sum_{j \neq i} \left(U_{\mathrm{FENE}}\left(r_{ij}\right) + U_{\mathrm{LJ}}\left(r_{ij}\right) + U_{\mathrm{C}}\left(r_{ij}\right)\right)\right] + F^E_{i},
\end{equation}
where $\nabla_i$ is the gradient with respect to the position of monomer $i$, and $r_{ij}$ is the distance between monomers $i$ and $j$. Note that in our MD model, both charge-charge and field-charge electrostatic forces are present and contribute to the net force acting on the charged monomers. 

\begin{figure}
    \centering
    \includegraphics[width=\linewidth]{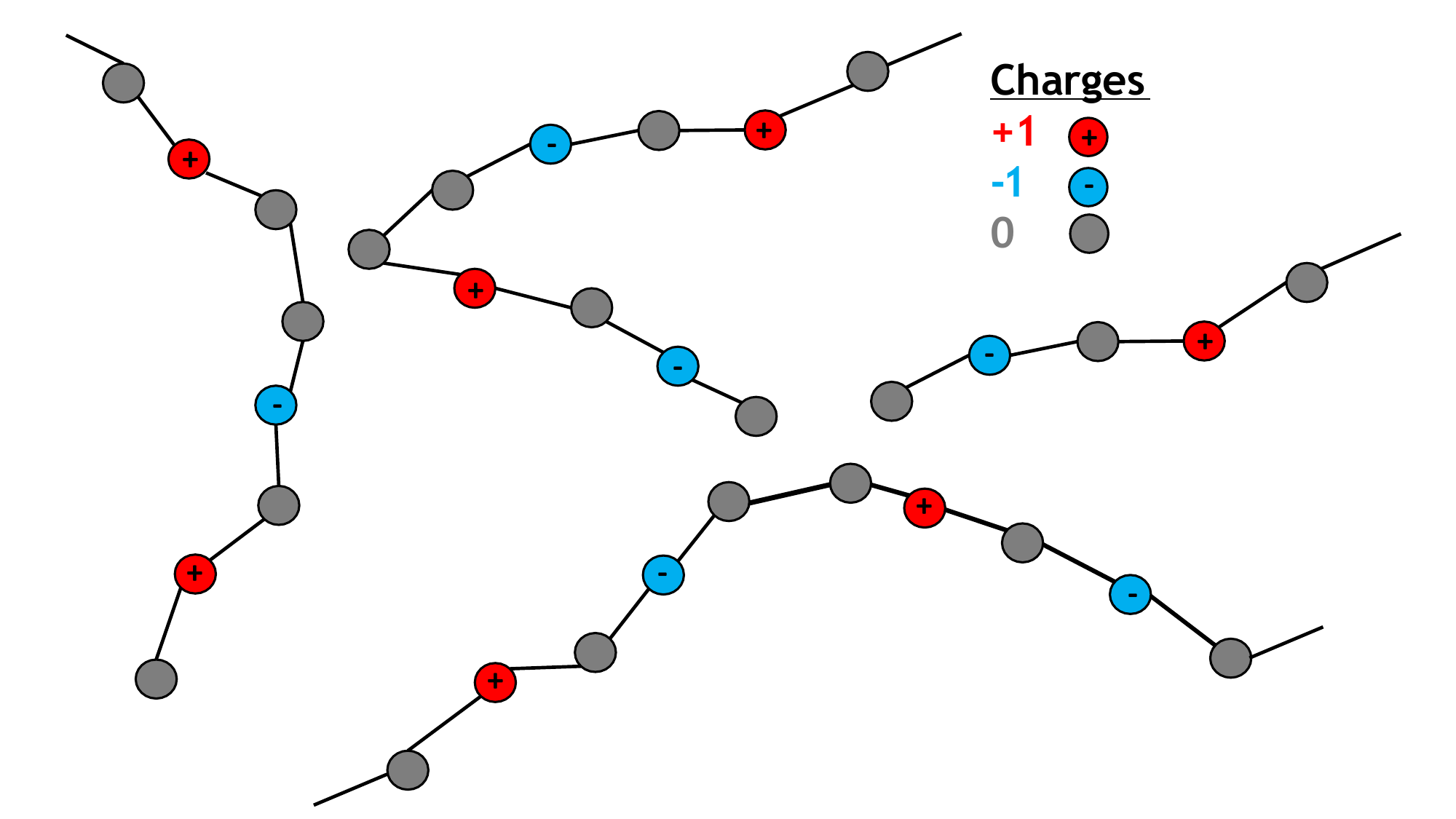}
    \caption{A visual representation of the linear polymer chain made using the Kremer-Grest model. The monomer coloring of red, blue, and gray represents a negatively charged, positively charged, and neutral monomer, respectively.}
    \label{fig:PolymerDesign}
\end{figure}

%We simulate 1000 polymers chains with length $N=50$ monomers in a cubic box with periodic boundary conditions. The charge sequence along the chain is fixed with a set charge pattern $(0,-1,0,1,0)$ every 5 monomers, which is designed to represent a fixed, repeating cosine series (see Fig. \ref{fig:PolymerDesign} and \ref{fig:PolymerCharge}). Additionally, this creates the effect of electric dipoles within the chain when neighboring charges are separated in an applied electric field.
We simulate 1000 polymers chains with length $N=50$ monomers in a cubic box with periodic boundary conditions. The charge sequence along the chain is fixed with a set charge pattern $(0,-1,0,1,0)$ every 5 monomers, which is designed to be cosine-like: a discrete, block realization of a net-zero, oscillating charge density representative of the periodic charge architecture found in many polyampholytes (see Fig. \ref{fig:PolymerDesign} and \ref{fig:PolymerCharge}). Additionally, this creates the effect of electric dipoles within the chain when neighboring charges are separated in an applied electric field.
Chains are initially placed at random in the box where we slowly increase the repulsion between monomers to push overlapping chains apart. An equilibration period of $10^4\tau$ is conducted in the NPT ensemble using the Nos\'{e}-Hoover thermostat and barostat, allowing the density of the melt to equilibrate at a temperature $T=1~k_BT/\epsilon$ and pressure $P=0~\sigma^3/\epsilon$. The damping parameters are set to $T_{damp}=100 \: \Delta\tau$ and $P_{damp}=1000 \: \Delta\tau$, where $\Delta\tau$ is the simulation timestep. We note that the equilibration period is approximately 2-3 times the longest Rouse relaxation time of the chains ($\tau_1\approx3860\tau$). After equilibration, the average box size and monomer density is measured from the last 10 frames spaced $100\tau$ apart to be $L\approx38.3\sigma$ and $\rho\approx0.89$, respectively. Prior to conducting the shear simulations, we switch to a constant volume (NVT) simulation for $100\tau$ to set the box dimensions. Shear flow simulations with applied electric fields, detailed below, are then performed in the NVT ensemble using the Nos\'{e}-Hoover thermostat with the same temperature settings as equilibration. Lastly, we set the timestep to $
\Delta\tau=0.005\tau$ for all simulations. For simplicity, all variables in the MD simulation are reported in LJ reduced units, $m=\sigma=\epsilon=k_B=1$. All units of time are reported in reduced LJ units of $\tau$.

After the above initial equilibration, we perform planar shear flow simulations with varying applied electric field strengths between $E=0-1~\epsilon/q_0\sigma $ in the $z$ direction orthogonal to the flow direction, $xz$. Specifically, for the non-equilibrium simulations we apply the SLLOD equations of motion\cite{evans1984nonlinear,todd2017nonequilibrium} to simulate planar shear with shear rates of $\dot{\gamma}=5e-4~\tau^{-1}$ and $\dot{\gamma}=1e-3~\tau^{-1}$ which equates to a Weissenberg number of $Wi\approx1.93$ and $Wi\approx3.86$, respectively. We perform the shear flow simulation for $15000\tau$, which is sufficient time to reach steady state at both shear rates and all applied electric field strengths.
% To ensure the system initializes and equilibrates properly, the NVT is held for the first $2,000,000$ time steps,  .
% All other boundaries of the flow are assumed to be rigid, with a slight attractive force between the walls and polymer chains to ensure that no-slip conditions are met.
% This attractive force is achieved by setting the Lennard-Jones potential cut-off distance to $r_c=1.4\sigma$ due to the positive potential gradient when $r>2^{1/6}\sigma$.
% Finally, when a pressure-driven flow scheme is used, the constant ``pressure differential'' is a fixed \texttt{addforce} (\texttt{addforce} is the relevant LAMMPS command) of $0.005$ on the monomers in the $x$ direction. 

\section{Viscoelasticity of Polymers with a Charge Density using the Rouse Model}

To elucidate the dependence of the electric field on the viscoelastic stress of an unentangled polymer melt with charges along the backbone, we modify the derivation by Rouse~\cite{rouse1953theory} following the approach outlined in chapter 5.3 of Doi's \textit{Introduction to Polymer Physics}~\cite{doi1996introduction}.
Consider a bead-spring polymer chain with $N$ beads connected by springs of stiffness $k$.
Additionally, each bead may be charged given by $Q_n$, where $n = 1, ..., N$, and the chain is in an electric potential, $\phi : \mathbb{R}^3 \mapsto \mathbb{R}$.
Let $\boldsymbol{R}_n \in \mathbb{R}^3$ be the position of bead $n$.
Then the potential energy of the chain is
\begin{equation} \label{eq:energy}
    \mathcal{U}\left(\boldsymbol{R}_1, ..., \boldsymbol{R}_{N}\right) = \sum_{n=1}^{N-1} \frac{k}{2}\left|\boldsymbol{R}_{n+1}-\boldsymbol{R}_{n}\right|^2 + \sum_{n=1}^{N} Q_n \phi\left(\boldsymbol{R}_n\right)
\end{equation}
The chain will prefer configurations of lower energy (\textit{i.e.}, beads with positive (negative) charges moving toward higher (lower) electric potential), but thermal fluctuations (e.g. Brownian motions) of the surrounding medium will cause the chain to tend towards configurations of higher entropy, and velocity gradients in the surrounding medium will cause beads in the chain to move with non-uniform velocity.
The interplay of these effects will give rise to an electro-viscoelasticity.
The overdamped Langevin equation (with velocity gradient in the surrounding flow) provides a formalism through which to model this interplay~\cite{doi1996introduction,balakrishnan2008elements,ottinger2012stochastic}:
\begin{equation} \label{eq:Langevin}
    \dot{\boldsymbol{R}}_n\left(t\right) = -\frac{1}{\zeta} \frac{\partial \mathcal{U}}{\partial \boldsymbol{R}_n}\biggr\rvert_{\boldsymbol{R}_1, ..., \boldsymbol{R}_{N}} + \boldsymbol{\nabla v}\cdot \boldsymbol{R}_n\left(t\right) + \boldsymbol{g}_n\left(t\right)
\end{equation}
where $\dot{\boldsymbol{R}} = \frac{\partial \boldsymbol{R}}{\partial t}$, $\zeta$ is a dampening coefficient (related to drag of the beads), $\boldsymbol{\nabla v}$ is the gradient of the velocity of the surrounding fluid, and $\boldsymbol{g}_n$ is the random force term arising from thermal fluctuations of the surrounding fluid.
The random force $\boldsymbol{g}_n$ is assumed to be independent of the chain configuration and satisfies the fluctuation-dissipation theorem\cite{callen1951irreversibility}; that is, it has zero mean $\langle{\boldsymbol{g}_n\left(t\right)}\rangle = \boldsymbol{0}$ and a variance written as,
\begin{subequations}
\begin{align}
    \langle{g_{i\alpha}\left(t\right) g_{j\beta}\left(t'\right)}\rangle &= 2 \delta_{\alpha \beta} \delta_{ij}\frac{k_BT}{\zeta} \delta({t}-{t'})
\end{align}
\end{subequations}
where $g_{i\alpha}$ is the $\alpha$ component of the random force on bead $i$.

Next, using equation \eqref{eq:energy}
\begin{equation} \label{eq:grad1}
    -\frac{1}{\zeta} \frac{\partial \mathcal{U}}{\partial \boldsymbol{R}_n}\biggr\rvert_{\boldsymbol{R}_1, ..., \boldsymbol{R}_{N}} = \frac{k}{\zeta} \left(\boldsymbol{R}_{n+1} - 2 \boldsymbol{R}_{n} + \boldsymbol{R}_{n-1}\right) + \frac{Q_n \boldsymbol{E}}{\zeta},
\end{equation}
where $\boldsymbol{E} = -\nabla\phi$.
If $\boldsymbol{E}$ is spatially varying, the resulting equations of motion are analytically intractable.
To make progress, we assume $\boldsymbol{E}$ is constant in space and time.
This effectively neglects electrostatic charge-charge interactions, but is a reasonable approximation when (1) the externally applied electric field (e.g., via a capacitor) is much greater in magnitude than fields due to charges, or (2) sufficient charge screening occurs.
Following the Rouse model derivation, it is convenient to treat $n \in \left[0, N\right]$ as a continuous variable such that $\bm{R}_n = \bm{R}\left(n, t\right)$ Recognizing that the term in the parentheses of \eqref{eq:grad1} is the form of a 2nd order central finite difference, we can rewrite \eqref{eq:Langevin} as
\begin{equation} \label{eq:dynam1}
    \dot{\boldsymbol{R}}\left(n, t\right) = \frac{k}{\zeta} \frac{\partial^2 \boldsymbol{R}}{\partial {n}^2} + \frac{\rho\left(n\right) \boldsymbol{E}}{\zeta} + \boldsymbol{\nabla v}\cdot \boldsymbol{R}\left(n, t\right) + \boldsymbol{g}\left(n, t\right)
\end{equation}
whereby, taking the continuum limit, $\rho : \left[0, N\right] \mapsto \mathbb{R}$ is the charge density along the backbone of the chain.
%To proceed, we integrate both sides of \eqref{eq:Langevin} with respect to $1 / N \int dn \cos\left(p \pi n / N\right)$; that is, we take a discrete cosine transform.
%Let $\boldsymbol{x}_p$ denote the $p$th mode, and assume the charge density can be represented by a cosine series such that
% \begin{align}
%     \boldsymbol{x}_p\left(t\right) &= \frac{1}{N} \int_0^N \cos\left(\frac{p \pi n}{N}\right) \boldsymbol{r} dn,\\
%     \rho\left(n\right) &= \sum_{i=1}^\infty {q}_i \cos \left(\frac{p \pi n}{N}\right) ,\\
%     \boldsymbol{g}_p\left(t\right) &= \frac{1}{N} \int_0^N \cos\left(\frac{p \pi n}{N}\right) \boldsymbol{g} dn.
% \end{align}
Using normalized coordinates, \eqref{eq:dynam1} can be written as,
\begin{equation}
    \dot{\boldsymbol{X}}_p = - \frac{\boldsymbol{X}_p}{\tau_p} + \frac{q_p \boldsymbol{E}}{\zeta_p}  + \boldsymbol{\nabla v}\cdot \boldsymbol{X}_p + \boldsymbol{g}_p \label{eq:Lag_Equations}
\end{equation}
where $\tau_p = \zeta_p / k_p = \tau_1 / p^2$ is the relaxation time of mode $p$, $\tau_1$ is the longest relaxation time, $q_p$ is the $p$th coefficient of the cosine series representation of the charge distribution, and
\begin{align}
    \zeta_p & = \begin{cases}
        N \zeta & p = 0 \\
        2 N \zeta & \text{otherwise}
    \end{cases} \\
    k_p &= \frac{2 \pi^2 k p^2}{N} = \frac{6 \pi^2 k_B T p^2}{N b^2} \\
\end{align}
are the drag and stiffness of mode $p$, respectively. From (\ref{eq:Lag_Equations}), one can observe an induced stretching of the Rouse mode as $-1/\tau_p(\bm{X}_p - \tilde{\bm{X}_p})$ where $\tilde{\bm{X}_p}=\tau_p q_p / \zeta _p \bm{E}$, adding additional stress from the applied electric field. 
% and,
% \begin{equation}
%     \overline{g_{p\alpha}\left(t\right) g_{r\beta}\left(t'\right)} = 2 \delta_{pr} \delta_{\alpha\beta} \frac{k_B T}{\zeta_p} \delta({t}-{t'}).
% \end{equation}

Using normalized coordinates, the deviatoric stress is written as, 
\begin{equation}
    \sigma_{\alpha\beta}=\frac{c}{N}\sum_{p}k_p\langle X_{p\alpha}X_{p\beta}\rangle
    \label{eq:avg_stress_Rouse}
\end{equation}
where $c$ is the number density of segments and $N$ is the number of segments in the chain. To demonstrate that this model is objective and anisotropic for the stress contributions from the electric field and flow directions while minimizing complexity of the derivation, we solve (\ref{eq:avg_stress_Rouse}) for a generalized dual shear flow in the $x$ and $y$ directions. We note that the 1D shear flow solution will not be able to show how the electric field interacts differently with the flow direction vs. the gradient direction and hence, we show the 2-dimensional flow case. 
% applying the relaxation time in terms of each $p$ mode, using $\zeta_p=2N\zeta$ for all modes $p\geq1$, and dropping higher order terms in $\gamma$ results in
This leads to the following equation (see Appendix~\ref{app:RouseDoi} for derivation)
\begin{equation}
    \begin{gathered}
        \sigma_{xy}=\frac{c}{N}\sum_{p=1}^\infty \frac{\zeta_p}{2}\Bigg(2\frac{\tau_p q_{p}^2E_xE_y}{\zeta_p^2}+\frac{k_BT}{k_p}(\dot{\gamma}_2+\dot{\gamma}_1)\\+2\frac{\tau_p^2}{\zeta_p^2}((q_{p}E_x)^2\dot{\gamma}_2+(q_{p}E_y)^2\dot{\gamma}_1)\Bigg)\\
        =\frac{c}{N}\sum_{p=1}^\infty \Bigg(\frac{\tau_1 q_{p}^2E_xE_y}{p^22N\zeta}+\frac{\tau_1 k_BT}{2p^2}(\dot{\gamma}_2+\dot{\gamma}_1)\\+\frac{\tau_1^2}{2Np^4\zeta}((q_{p}E_x)^2\dot{\gamma}_2+(q_{p}E_y)^2\dot{\gamma}_1)\Bigg)
    \end{gathered}
    \label{eq:Rouse_dualshear_allmodes}
\end{equation}
where $\dot{\gamma}_1 =\partial v_x/\partial y$ and $\dot{\gamma}_1 =\partial v_y/\partial x$ are the applied shear rates in the $x$ and $y$ directions. The third term in the sum of (\ref{eq:Rouse_dualshear_allmodes}) demonstrates there is a directional dependence of the electric field with the flow direction. We also observe that the stress contribution from the electric field scales with the relaxation times of each Rouse mode, $\sigma_{xy} \sim (\tau_1/p^2)^2 q_pE_y^2 \dot{\gamma} $ due to the charges being connected along the chain. However, for a defined charge sequence, we expect that the contribution of the stress from the charge sequence will be dependent on the spacing of the charges along the chain and therefore only contribute to the stress for a specific Rouse mode. We demonstrate this by simplifying the above expression for a defined charge sequence.

\begin{figure}
    \centering
    \includegraphics[width=\linewidth]{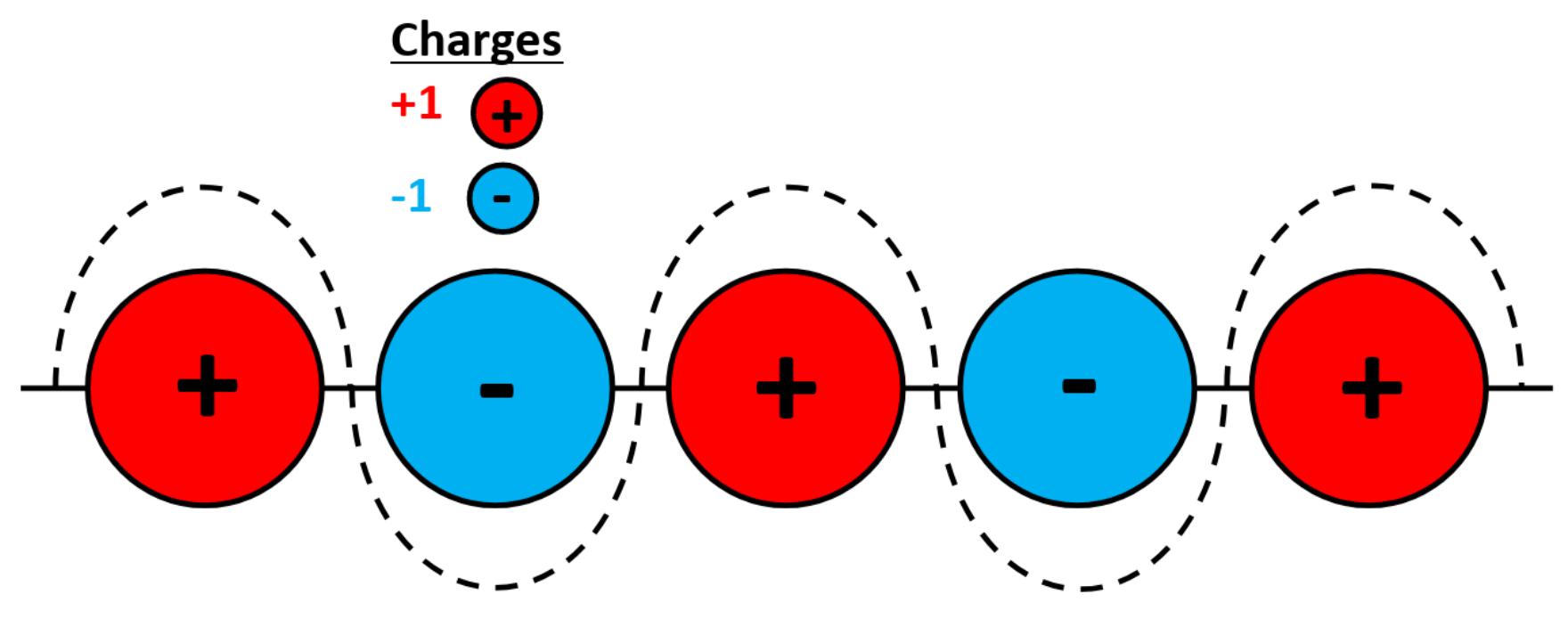}
    \caption{An example visual representation of a charge sequence along a chain approximated by an oscillatory function (The dashed line along the chain). The monomer coloring of red and blue represents a positively charged monomer and a negatively charged monomer, respectively.}
    \label{fig:PolymerCharge}
\end{figure}

When considering a specific charge sequence with both positively and negatively charged monomers spaced along the chain, a smooth cosine representation is prescribed as an approximation of the oscillatory charge sequence of monomers. 
An example case of this is shown in Fig. \ref{fig:PolymerCharge}, where the cosine function is a smooth approximation of the charge sequence along the chain that would stretch under an electric field in a dipole-like nature.
In general, one can represent a charge sequence through an infinite Fourier series representation if a more complex charge sequence is used.
However, for simplicity, we only consider a charge density with a single non-zero mode $p=f$ such that $q_p=q\delta(p-f)$. Then, the stress contribution from this mode can be written as, 
\begin{equation}
    \begin{gathered}
        \sigma_{xy}^{p=f}=\frac{c}{N}\Bigg(\frac{\tau_1 q^2 E_{x}E_{y}}{f^22N\zeta}+\frac{\tau_1 k_BT}{2f^2}(\dot{\gamma}_2+\dot{\gamma}_1)\\+\frac{q^2\tau_1^2}{2Nf^4\zeta}({E_{x}}^2\dot{\gamma}_2+{E_{y}}^2\dot{\gamma}_1)\Bigg)
    \end{gathered}
\end{equation}
By applying Rouse's definition of the relaxation time, $\tau_R=\tau_1=\frac{\zeta N^2b^2}{3\pi^2k_bT}$ (which assumes the chains are Gaussian in each Rouse segment), and summing over all other modes for the term $\frac{\tau_1 k_BT}{2p^2}$ (which is not dependent on charge density and is therefore not restricted to only mode f), a final form of the shear stress in terms of simple material quantities can be found as
\begin{equation}
    \begin{gathered}
        \sigma_{xy}=\epsilon_{dielec}E_{x}E_{y}+\eta(\dot{\gamma}_2+\dot{\gamma}_1)+\tau_f\epsilon_{dielec}({E_x}^2\dot{\gamma}_2+{E_{y}}^2\dot{\gamma}_1)
    \end{gathered}
    \label{eq:simplified_Rouse}
\end{equation}
where $\tau_f=\tau_1/f^2$ is the relaxation time of mode $f$, $\eta=\frac{c\zeta Nb^2}{36}$ is the viscosity with no electric field, and $\epsilon_{dielec}=\frac{c\tau_fq^2}{4N^2\zeta}=\frac{cq^2b^2}{6\pi^2f^2k_BT}$ is the dielectric permittivity with $q$ as the charge amplitude of the cosine series, $b$ as the Kuhn length, and $k_BT$ as the thermal energy.
Again, from both the single mode contribution in (\ref{eq:simplified_Rouse}) and total stress from an arbitrary sequence in (\ref{eq:Rouse_dualshear_allmodes}), one can see how the strain rate affects the viscosity without an applied electric field, but leaves an anisotropic dependence of the electric field on the scaling of the viscosity with strain rate. 
Notably, \emph{this directional dependence of $\tau_f\epsilon_{dielec}({E_x}^2\dot{\gamma}_2+{E_{y}}^2\dot{\gamma}_1)$ requires careful consideration when upscaling to the continuum model derivation}.

A formal derivation of continuum-level constitutive equations from the Rouse model, analogous to those for the original Rouse and Zimm models,\cite{lodge1971constitutive} is left for future work.
Instead, in the following section we propose a minimal continuum model for the stress response, inspired by the Rouse-level description and by derivations of the upper-convected Maxwell model from oppositely charged harmonic dumbbells (see Appendix~\ref{app:Dumbbell}). We then discuss the physical implications and limitations of the model before comparing with particle-based simulation results. 

\section{Electro-viscoelasticity of a polymer melt in the continuum limit}

\subsection{Classical constitutive models for electrorheological fluids}

As described in  Refs. \citenum{ruzicka_electrorheological_2000,prusa_flow_2012,ruzicka_modeling_2004}, one can model an electrorheological fluid using the stress tensor,
\begin{equation}
    \sigma_{ij}=2\eta_0D_{ij}+\eta_1(D_{ik}E_kE_j+D_{jk}E_kE_i)
    \label{eq:classical_ER_stress}
\end{equation}
where $D_{ij}$ is the deformation rate tensor and $E_iE_j$ is the dyadic of the electric field with itself.
With the term $\eta_1(D_{ik}E_kE_j+D_{jk}E_kE_i)$, the overall stress is dependent on the coupling between the shear rates and electric field to arrive at an overall viscosity that scales as $E^2$.
% However, considering the Rouse model, issues begin to arise with consistency.
The principal issue when describing the stress response of polymers, as seen from the chain-level Rouse description, is that one would need to use the velocity gradient instead of the symmetrized velocity gradient to enforce the directional dependence of the increase in viscosity that the polymer experiences under an electric field during flow.
Purely using the velocity gradient in (\ref{eq:classical_ER_stress}) for the constitutive statement is not an objective measure of stress, as rigid rotation will cause a different measure of viscosity of the fluid.
Using this form also predicts that a Couette flow will have the viscosity scaling as $(E_1^2+E_2^2)$, meaning any rotation of the electric field about the $(x_1-x_2)$ plane will produce the same shear stress response. 
%With this, the typical linear constitutive responses will not be the correct approach to arrive at the desired properties.
Hence, existing continuum approaches cannot recover key physics observed in the more detailed Rouse model.

\subsection{Upper-convected Electro-Maxwell Model}
To recover the correct scaling at the continuum limit, we step back to consider viscoelastic models more broadly and note the general applicability of Maxwell fluids for modeling polymers. 
The constitutive equation from the Rouse model assumes that the total polymer stress $(\bm{\sigma})$ is a sum of the stress from individual Rouse modes $(\bm{\sigma}^i)$,
\begin{equation}
    \bm{\sigma}=\sum_i^p \bm{\sigma}^i
\end{equation}
where each mode obeys an upper-convected Maxwell (UCM) model,
\begin{equation}
    \stackrel{\nabla}{\bm{\sigma}^i}+\lambda^i\bm{\sigma}^i = 2\eta^i\bm{D}
    \label{eq:UCM_form}
\end{equation}
where $\lambda^i$ is the relaxation time of mode $i$ to be consistent with standard notation of the characteristic relaxation time used in the UCM model. Furthermore, as mentioned for the Rouse model, the polymer itself is experiencing an internal stretch and alignment due to the electric field and dipole interaction.
To capture this in the continuum limit, we derive the upper-convected Maxwell equation for an oppositely-charged harmonic dumbbell model as a first approximation (see Appendix B).

%To derive the continuum model, one can decompose the total stress into the electrostatic Maxwell stresses from polarization ($E_iP_j-\frac{1}{2}E_kP_k\delta_{ij}$), polymer stresses ($\sigma_{ij}^{p}$), and pressure $(-p\delta_{ij})$, such that
To derive the continuum model, one can decompose the total stress into the Maxwell stress, polymer stresses, and pressure.
The Maxwell stress is constructed from the dyadic of the electric field with the displacement field $D_i = \epsilon_0 E_i + P_i$. 
Decomposing this dyadic, $E_i D_j = \epsilon_0 E_i E_j + E_i P_j,$
separates a vacuum contribution from a polarization contribution. 
The first term is present with or without the polymer and is constant for the spatially uniform applied field considered here. In an incompressible melt, this term, together with the isotropic contributions, exerts no material stress and is absorbed into the pressure. 
The rheologically relevant contribution is therefore $E_iP_j$, where $P_j = \epsilon_{dielec} E_j$ is the polarization. In the context of this continuum framework, $\epsilon_{dielec}$ represents the effective polarization susceptibility of the chain, $\epsilon_0 \chi_p$, and can be directly related to the bulk dielectric constant of the melt, $\epsilon_r$, through the standard relation $\epsilon_r = 1 + 4\pi\chi_p$.
Upon decomposition:
\begin{equation}
    \sigma_{ij}=\sigma_{ij}^{p}+\epsilon_{dielec}(E_iE_j-\frac{1}{2}E_kE_k\delta_{ij})-p\delta_{ij}.
\end{equation}
%where $P_i=\epsilon_{dielec}E_i$ is the polarization vector, and $\epsilon_{dielec}$ is the apparent dielectric permittivity of the Rouse chain from polarization.
Note that, in general, $\epsilon_{dielec}$ may be coupled with multiple Rouse modes, as shown in \ref{eq:simplified_Rouse}, but is simplified here to a singular material function which captures all charge frequency mode contributions. 
Taking the initial total stress decomposition allows for the polymer contributions (the viscoelastic contributions to the system) to be completely excluded from the Maxwell elastostatic stresses (which capture the total momentum conservation of a moving dielectric medium \cite{griffiths2017introduction}), and will provide a convenient notation when checking for compliance with the second law of thermodynamics.
Next, we can describe to a first-order approximation the total polymer stress using the UCM model having the form shown in (\ref{eq:UCM_form}). This is the simplest model for polymer melts that captures the first normal stress difference (from the finite rotation of polymers under flow \cite{macosko1994rheology}) and transient viscoelastic behavior. However, it assumes a single characteristic relaxation time $\lambda$ and, as mentioned previously, fails to capture shear thinning at higher shear rates. While a more complex model could be derived from the ground up using non-equilibrium thermodynamic formalisms (\textit{e.g.}, GENERIC\cite{grmela1997dynamics}), we start with the simplest case first to demonstrate the decoupling of the relaxation timescales contributing to the stress from both flow and applied electric fields. To this end, we modify the UCM equation with an electric-field coupling term,
\begin{equation}
    \begin{gathered}
        \sigma_{ij}^{p}+\lambda \stackrel{\nabla}{\sigma^{p}_{ij}}+\lambda_f\epsilon_{dielec}\stackrel{\nabla}{(E_iE_j)}=2\eta D_{ij} \label{eq:ElectroMaxwellModel}
    \end{gathered}
\end{equation}
where $\stackrel{\nabla}{\sigma_{ij}}=\dot{\sigma}_{ij}+v_k\sigma_{ij,k}-v_{i,k}\sigma_{kj}-\sigma_{ik}v_{j,k}$ is the upper convected derivative, $\lambda$ is the apparent relaxation time, $\lambda_f$ is an effective relaxation time for the frequency of the charge density, and $\eta$ is the shear viscosity without an electric field present.
When no electric field is applied, the original UCM equation is recovered.

%We make two physical assumptions in the above model. Similar to the UCM model, we assume a single dominant relaxation time is coupled to the evolution of the polymer stress. The second assumption is that there is a separate effective relaxation time $\tau_f$ that couples to the evolution of stress from the electric field. The latter is based on the Rouse model derivation, which shows that for a defined charge sequence, the stress contribution from the electric field will be coupled to a single Rouse mode with relaxation time of $\tau/f^2$. 
There are two assumptions implicit via the form of the newly developed model: (1) there is a single dominant relaxation time coupled to the stress evolution, and (2) there is a distinct relaxation time, $\lambda_f$, that couples the evolution of the stress from the electric field.
While the latter is inspired from the Rouse model, we emphasize that $\epsilon_{dielec}$ and $\lambda_f$ are not taken to be fundamental linear response constants (in general), but instead \emph{apparent material functions}.
The UCEM equation is a minimal continuum structure whose purpose is to reproduce the directional, field-coupled stress established at the Rouse level; the coefficients $\epsilon_{dielec}$ and $\lambda_f$ compress into two scalar functions the several distinct contributions retained in the MD system (finite chain extensibility (FENE), Coulomb interactions) that the analytical derivation cannot resolve separately.
This interpretation is similar to a White-Metzner approach, where the upper-convected Maxwell model assumes the viscosity and relaxation time are state-dependent on the shear rate \cite{tseng_revisitation_2021} to capture shear-thinning and low relaxation times at high shear rates.
Their dependence on field stress is expected rather than anomalous.
We later extract $\epsilon_{dielec}\left(E^2\right)$ and $\lambda_f\left(E^2\right)$ as state dependent material functions, dependent on $E^2$, from MD.
Because the functions depend on the field invariant, $E^2$, rather than $E$ itself, their dependence is frame-indifferent by construction.
Further the MD simulations demonstrate that the UCEM captures the correct tensorial structure of the electro-rheological response, while the magnitude of that response is carried by material functions that must be measured rather than assumed constant.

Several physical justifications for these assumptions are warranted. First, the upper-convected derivative $\stackrel{\nabla}{(E_i E_j)}$ is necessary because electric polarization in the charged polymers considered herein is coupled to local chain deformation, where $E_i E_j$ provides the direction and force amplitude of the chain polarization. Although it may seem counterintuitive to apply a convected derivative to the electric field rather than to a polymer microstructural quantity, polymer line elements are not explicitly represented at the continuum scale, whereas the field is. Moreover, only the relative orientation between the field and the polymer matters; reorientation of the polymer in a fixed field is physically equivalent to reorientation of the field relative to a fixed polymer. As shown in Appendix~\ref{app:Dumbbell}, this structure arises naturally for an elastic dumbbell with a dipole-like structure (equal but opposite charges on each mass end), which produces the same $\stackrel{\nabla}{(E_i E_j)}$ contribution. One can further verify that the interaction is objective in general, as $E_i^*E_j^*=Q_{ik}E_kE_pQ_{jp}$, where $Q_{ik}$ are the components of an orthogonal rotation tensor.

Secondly, the applied electric field induces chain stretching along field lines through the oscillating charge sequence, creating quasi-one-dimensional polarized regions embedded in the polymer backbone. These polarized segments rotate and stretch with the flow like material line elements, requiring an objective convected derivative to properly describe their evolution. As with stress induced by chain stretching and rotation in flow, the standard material derivative is insufficient for capturing the objective response of viscoelastic fluids, whose behavior is intrinsically tied to deformation history. Standard formulations lacking this derivative implicitly assume that polarization is spatially fixed or responds instantaneously, neglecting the finite charge redistribution time. (The UCEM form also simplifies the relaxation of polarization stresses, as the time-dependent response under an electric field depends on both the transient end-to-end vector and the configuration tensor in the elastic dumbbell model.) This explains why the viscosity enhancement scales generally as $\lambda_f \epsilon_{dielec} E^2$ rather than $\lambda \epsilon_{dielec} E^2$: the electric field drives rearrangements at the length scale of the charge sequence (mode $f$), not the global chain length.

We next compare the stress contribution from the electric field to that of the Rouse model. To do so, we first define a velocity gradient for the dual shear flow, where
\begin{equation}
    v_{i,j}=\begin{bmatrix}
        0 & \dot{\gamma}_1  & 0 \\
        \dot{\gamma}_2 & 0 & 0\\
        0& 0 & 0
    \end{bmatrix}
\end{equation}
which is symmetrized for the deformation rate tensor,
\begin{equation}
     D_{ij}=\frac{1}{2}\begin{bmatrix}
        0 & \dot{\gamma}_1 +  \dot{\gamma}_2 & 0 \\
        \dot{\gamma}_2 + \dot{\gamma}_1 & 0 & 0\\
        0& 0 & 0
    \end{bmatrix}
\end{equation}
Then, when applying the definition to the upper-convected derivative of the electric field dyadic under steady-state assumptions, one can show
\begin{equation}
    \begin{gathered}
        \stackrel{\nabla}{(E_iE_j)}=-v_{k,i}E_kE_j-E_iE_kv_{k,j}
    \\=-\begin{bmatrix}
        2E_1E_2\dot{\gamma}_1 & E_2^2\dot{\gamma}_1+E_1^2\dot{\gamma}_2 & 0 \\ E_2^2\dot{\gamma}_1+E_1^2\dot{\gamma}_2 & 2E_1E_2\dot{\gamma}_2 & 0 \\ 0&0&0
    \end{bmatrix}
    \end{gathered}
\end{equation}
where the coupling term from the Rouse model is recovered for the two-dimensional, dual shear flow example. 
This means that, to capture the same behavior observed in the Rouse model, the upper-convected time derivative yields the correct coupling to the shear rate as opposed to using other objective material derivatives (such as the Jaumann or Green-Naghdi).

To further simplify the model, we assume the polymer to be incompressible, i.e.,
\begin{equation}
    v_{k,k}=D_{kk}=0
\end{equation}
where repeated indices imply summation and where commas in the subscript denote a partial derivative with respect to spatial coordinate (i.e., $\alpha_{,k} = \partial \alpha / \partial x_k$).
Finally, to ensure linear momentum conservation, the force balance with no electric field gradients is kept in the case of a Stokes flow
\begin{equation}
    \sigma_{ij,j}+\rho f_i=p_{,i}
\end{equation}
where $f$ is an arbitrary volume force, $\rho$ is the polymer density, and $p$ is the pressure field.
For the simplest case of non-relativistic, steady-state electric fields (the electric field is also assumed to be applied as a step function in some later sections when examining the transient viscoelastic behavior of the polymer), they must satisfy the following conditions
\begin{equation}
    E_{i,i}=0
\end{equation}
\begin{equation}
    \epsilon_{ijk}E_{j,k}=0
\end{equation}
which are trivially satisfied when the electric field is homogeneous.  

\subsection{Thermodynamic consistency analysis}
To verify that the UCEM model is thermodynamically consistent, we show that positive energy dissipation is guaranteed under suitable conditions using the Clausius-Duhem (dissipation) inequality. Starting from the Helmholtz free energy in terms of the polymer and electric energy storage,
\begin{equation}
    \Psi = \frac{G}{2}(B_{ii}-3)+\frac{\epsilon_{dielec}}{2}E_iE_i
\end{equation}
where $G=\frac{\eta}{\lambda}$ and $\epsilon_{dielec}$ are positive definite to ensure a positive energy density of the medium, and $B_{ij}=\frac{\sigma_{ij}^p}{G}+\delta_{ij}$ is the left Cauchy-Green tensor \cite{FUSI20101263}.
The material time derivative of this free energy storage can then be found as
\begin{equation}
    \frac{D(\Psi)}{Dt} = GB_{ik}D_{ki}-\frac{G}{2\tau}(B_{ii}-3)+\epsilon_{dielec}E_i\frac{D({E}_i)}{Dt}
\end{equation}
For an incompressible, isothermal flow, the dissipation inequality is defined as
\begin{equation}
    \begin{gathered}
        \Phi=\sigma_{ij}D_{ij}- \frac{D(\Psi)}{Dt}\geq 0\\
        \sigma_{ij}^{p}D_{ij}+\epsilon_{dielec}E_iE_jD_{ij}- \frac{D(\Psi)}{Dt} \geq 0
    \end{gathered}
\end{equation}
where the pressure and $E_kE_k$ terms are zero as $D_{kk}=0$.
By rewriting the form of the strain rate tensor in the polymer constitutive response and substituting, one can find
\begin{equation}
    \begin{gathered}
        (2\eta D_{ij}-\lambda \stackrel{\nabla}{\sigma^{p}_{ij}}-\lambda_f\epsilon_{dielec}\stackrel{\nabla}{(E_iE_j)})D_{ij}+\epsilon_{dielec}E_iE_jD_{ij}\\- GB_{ik}D_{ki}+\frac{G}{2\lambda}(B_{ii}-3)-\epsilon_{dielec}E_i\frac{D({E}_i)}{Dt} \geq 0
    \end{gathered}
\end{equation}
where $2\eta D_{ij}D_{ij}$ is the only positive-definite term, meaning that $\eta\geq0$ for the irreversible dissipation.
The new term of interest for the dissipation function is 
\begin{equation}
    \lambda_f\epsilon_{dielec}\stackrel{\nabla}{(E_iE_j)}D_{ij}
\end{equation}
In this term, the electric field acts upon a distinct relaxation mode of the chain ($f$); a negative relaxation time ($\lambda_f < 0$) would imply non-physical energy generation via negative damping, as the term facilitates the exchange of energy between the external electric field and the polymer's internal free energy functions.
Thus, a sufficient condition for the dissipation function to be non-negative is $\lambda_f\geq0$. This condition is automatically satisfied from the definition of the Rouse mode relaxation, $\tau_f$ which is determined by the charge sequence mode $f$ and is strictly positive.

\subsection{Transient polarization response to a step electric field}
To determine the steady-state properties of the system under flow, we apply the electric field as a step function within the MD simulations. Given that chain polarization is governed by a specific relaxation time $\lambda_f$, it is necessary to investigate how this transient response diverges from the behavior predicted by the standard Maxwell stress tensors, which assume an instantaneous stress jump. To analyze this, we consider a quasi-static medium subjected to a step-electric field defined as $E_2 = E_0 H(t)$, where $H(t)$ denotes the Heaviside step function.
When applying this to the constitutive response, one can find the stress response to be simplified to the following equations,
\begin{equation}
    \begin{gathered}
            \sigma_{11}=\sigma_{33}=-\frac{1}{2}\epsilon_{dielec}E_0^2-p\\
            \sigma_{22}=\sigma_{22}^p+\frac{1}{2}\epsilon_{dielec}E_0^2-p\\
            \sigma_{22}^p + \lambda\dot{\sigma}_{22}^p=-\lambda_f\epsilon_{dielec}E_0^2\delta(t)
    \end{gathered}
\end{equation}
where $\delta(t)$ is the delta function.
For $t \geq 0$, the solution for the polarization stress component $\sigma_{22}^p$ following the step-excitation $H(t)$ is derived as,
\begin{equation}
    \sigma_{22}^p(t)=-\frac{\lambda_f}{\lambda}\epsilon_{dielec}E_0^2e^{-t/\lambda}. \label{eq:Transient_s_22_p}
\end{equation}
Consequently, the total normal stress $\sigma_{22}$ is expressed as,
\begin{equation}
    \sigma_{22}(t)=\epsilon_{dielec}E_0^2\Bigg(\frac{1}{2}-\frac{\lambda_f}{\lambda}e^{-t/\lambda}\Bigg)-p.
\end{equation}
By evaluating the second normal stress difference, $N_2 = \sigma_{22} - \sigma_{33}$, we can eliminate the isotropic background pressure term to obtain,
\begin{equation}
    N_2=\sigma_{22}-\sigma_{33}=\epsilon_{dielec}E_0^2\Big(1-\frac{\lambda_f}{\lambda}e^{-t/\lambda}\Big), \label{eq:N2_Transient_E}
\end{equation}
This relationship indicates that the medium's polarization undergoes an instantaneous initial jump to $(1 - \frac{\lambda_f}{\lambda})$ and approaches its full steady-state polarization approximately $5\lambda$ after the field application. This also further predicts that in the limit where $\lambda_f \ll \lambda$, the polarization immediately reaches its steady-state value at $t = 0$. Thus, the magnitude of the initial elastic jump serves as a direct metric for how significantly the charge sequence dynamics influence the overall chain response.
Furthermore, at the moment of peak polarization stress amplitude ($t = 0$), the Helmholtz free energy density is given by
\begin{equation}
    \Psi = \frac{\epsilon_{dielec}E_0^2}{2}\Big(1-\frac{\lambda_f}{\lambda} \Big),
\end{equation}
where $\lambda_f > \lambda$, the model would predict a non-physical negative energy density, implying the polymer is generating more energy than is supplied by the external electric field.
Thus, stable solutions require the charge relaxation time to satisfy $0 \leq \lambda_f \leq \lambda$. In the Rouse model, this corresponds to the constraint that the frequency of the charge sequence cannot exceed the number of beads.

\subsection{Couette flow}

One can solve for a thin, steady-state, simple shear with a uniform electric field, starting with the following kinematic definitions
\begin{equation}
    v_i(x_1,x_2,x_3) = 
    \begin{bmatrix}
        \dot{\gamma} x_2\\0\\0
    \end{bmatrix},
\end{equation}
and
\begin{equation}
    v_{i,j} = 
    \begin{bmatrix}
        0&\dot{\gamma}&0\\0&0&0\\0&0&0
    \end{bmatrix}.
\end{equation}
One can then find the upper-convected derivatives as
\begin{equation}
   \stackrel{\nabla}{\sigma_{ij}^p}=-v_{i,k}\sigma_{k,j}^p-\sigma_{ik}^pv_{j,k}
    =-\dot{\gamma}\begin{bmatrix}
        2\sigma_{12}^p & \sigma_{22}^p &0\\ \sigma_{22}^p & 0 &0\\0&0&0
    \end{bmatrix},
\end{equation}
and
\begin{equation}
   \stackrel{\nabla}{(E_iE_j)}=-v_{i,k}E_kE_j-E_iE_kv_{j,k}
    =-\dot{\gamma}\begin{bmatrix}
        2E_1E_2 & E_2E_2 & 0 \\ E_2E_2 & 0 & 0 \\ 0&0&0
    \end{bmatrix}.
\end{equation}
From these equations, the electro-Maxwell model in matrix form can be written as,
\begin{equation}
    \begin{gathered}
    \begin{bmatrix}
        \sigma_{11}^{p} &\sigma_{12}^{p}&\sigma_{13}^{p}\\\sigma_{12}^{p}&\sigma_{22}^{p}&\sigma_{23}^{p}\\ \sigma_{13}^{p} & \sigma_{23}^{p} & \sigma_{33}^{p}
    \end{bmatrix}
    -\lambda\dot{\gamma}\begin{bmatrix}
        2\sigma_{12}^{p} & \sigma_{22}^{p} & 0 \\ \sigma_{22}^{p} & 0 &0 \\ 0& 0& 0
    \end{bmatrix}
    \\-\lambda_f\epsilon_{dielec}\dot{\gamma}\begin{bmatrix}
        2E_1E_2 & E_2E_2 &0\\ E_2E_2 & 0 &0\\0&0&0
    \end{bmatrix} =
    \eta\begin{bmatrix}
        0 &\dot{\gamma} &0\\\dot{\gamma}&0 &0 \\ 0&0&0
    \end{bmatrix}
    \end{gathered}
\end{equation}
where the polymer stresses are
\begin{equation}
    \begin{gathered}
        \sigma_{13}^{p}=\sigma_{23}^{p}=\sigma_{22}^{p}=\sigma_{33}^{p}=0,\\
        \sigma_{12}^{p}=(\lambda_f\epsilon_{dielec}E_2^2+\eta)\dot{\gamma},\\
        \sigma_{11}^{p}=2\lambda(\lambda_f\epsilon_{dielec}E_2^2+\eta)\dot{\gamma}^2+2\lambda_f\epsilon_{dielec}E_1E_2\dot{\gamma}.
    \end{gathered}
\end{equation}
Each non-zero component of the stress now can be found in the form 
\begin{equation}
    \begin{gathered}
        \sigma_{11}=2\lambda(\lambda_f\epsilon_{dielec}E_2^2+\eta)\dot{\gamma}^2+2\lambda_f\epsilon_{dielec}E_1E_2\dot{\gamma}\\-p+\frac{\epsilon_{dielec}}{2}(E_1^2-E_2^2-E_3^2),\\
        \sigma_{12}=(\lambda_f\epsilon_{dielec}E_2^2+\eta)\dot{\gamma} + \epsilon_{dielec}E_1E_2,\\
        \sigma_{22}=\frac{\epsilon_{dielec}}{2}(E_2^2-E_1^2-E_3^2)-p,\\
        \sigma_{33}=\frac{\epsilon_{dielec}}{2}(E_3^2-E_1^2-E_2^2)-p,\\
        \sigma_{13}=\epsilon_{dielec}E_1E_3,\\
        \sigma_{23}=\epsilon_{dielec}E_2E_3.
    \end{gathered}
\end{equation}
With this, the shear stress from the upper-convected electro-Maxwell model at a constant shear rate matches the stress response observed in the Rouse model. Specifically, the upper-convected derivative describes a fluid in which a material element stretches and rotates during flow \cite{macosko1994rheology}.
Thus, the increase in viscosity of a polymer chain under a constant electric field and shear rate is due to the polarization of $\sigma_{22}^p$ as the chain is reoriented by the flow, shifting it from the polarization direction toward alignment with the flow.

Furthermore, one can estimate the distinct relaxation timescales ($
\lambda$ and $\lambda_f$), and the dielectric permittivity from the fluid's total stresses under an $E_2$ electric field, by examining the polarization at a constant shear rate as,
\begin{equation}
    \sigma_{22}(E_2)-\sigma_{33}(E_2)=\epsilon_{dielec}E_2^2, \label{eq:Polar1}
\end{equation}
\begin{equation}
    \sigma_{12}(E_2)-\sigma_{12}(E_2=0)=\lambda_f\epsilon_{dielec}E_2^2\dot{\gamma},\label{eq:Polar2}
\end{equation}
\begin{equation}
    \sigma_{11}(E_2)-\sigma_{33}(E_2)=2\lambda\lambda_f\epsilon_{dielec}E_2^2\dot{\gamma}^2+2\lambda\eta\dot{\gamma}^2=2\lambda\sigma_{12}(E_2)\dot{\gamma},\label{eq:Polar3}
\end{equation}

% \begin{figure}
%     \centering
%     \includegraphics[width=\linewidth]{Pictures/Bestfit_Long_0.0001.pdf}
%     \caption{Polarization stress for a constant shear rate of $\dot{\gamma}=0.0001$. The best fit values for the $\sigma_{22}(E_2)-\sigma_{33}(E_2)$, $\sigma_{12}(E_2)-\sigma_{12}(E_2=0)$, and $\sigma_{11}(E_2)-\sigma_{33}(E_2)$ polarizations are shown, respectively, relative to the quadratic scaling with the electric field.}
%     \label{fig:CouettePolarLong0.0001}
% \end{figure}
% \begin{figure}
%     \centering
%     \includegraphics[width=\linewidth]{Pictures/Bestfit_Long_0.001.pdf}
%     \caption{Polarization stress for a constant shear rate of $\dot{\gamma}=0.001$. The best fit values for the $\sigma_{22}(E_2)-\sigma_{33}(E_2)$, $\sigma_{12}(E_2)-\sigma_{12}(E_2=0)$, and $\sigma_{11}(E_2)-\sigma_{33}(E_2)$ polarizations are shown, respectively, relative to the quadratic scaling with the electric field.}
%     \label{fig:CouettePolarShort0.001}
% \end{figure}

\begin{figure}
    \centering
        \sidesubfloat[]{\includegraphics[width=.8\linewidth]{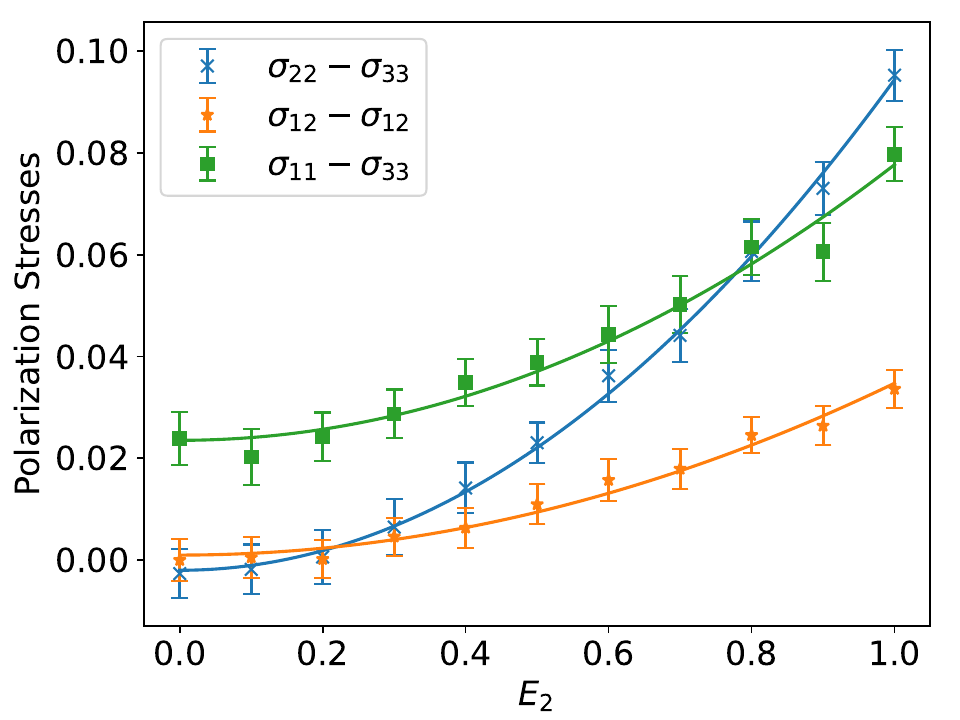}}\\
        \sidesubfloat[]{\includegraphics[width=.8\linewidth]{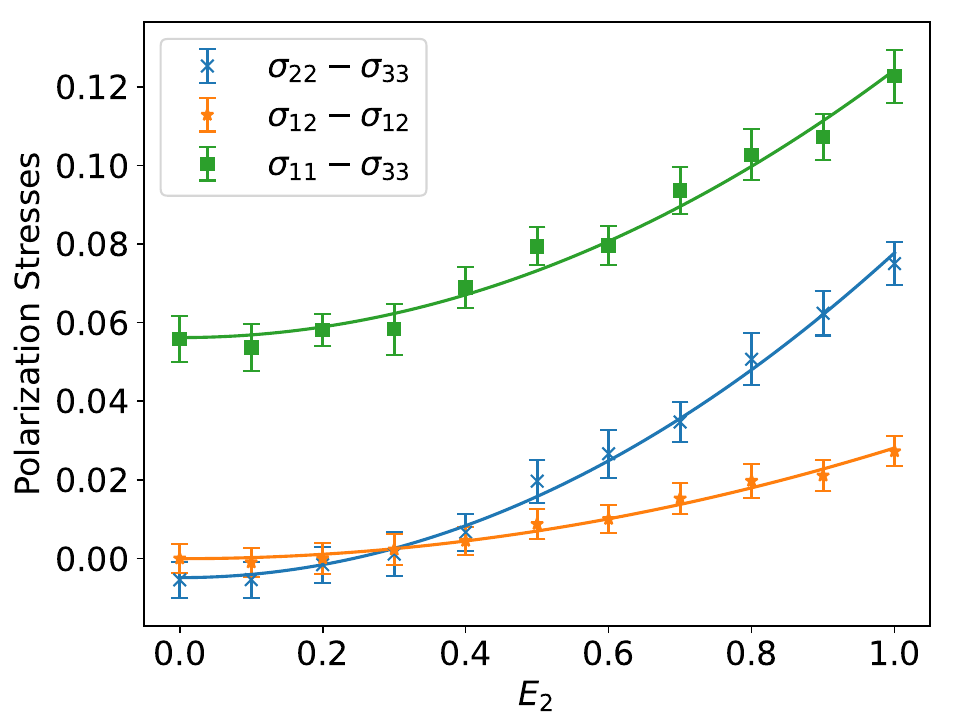}}
    \caption{Polarization stress for a constant shear rate of (a) $\dot{\gamma}=5e-4$ and (b) $\dot{\gamma}=1e-3$. The best fit curves for the $\sigma_{22}(E_2)-\sigma_{33}(E_2)$, $\sigma_{12}(E_2)-\sigma_{12}(E_2=0)$, and $\sigma_{11}(E_2)-\sigma_{33}(E_2)$ polarizations are shown, respectively, relative to the quadratic scaling with the electric field. Error bars represent one standard deviation from the mean.}
    \label{fig:CouettePolarFits}
\end{figure}

% \begin{figure}
%     \centering
%     \includegraphics[width=\linewidth]{Pictures/Dielec.pdf}
%     \caption{The dielectric permittivity calculated using \eqref{eq:Polar1} at the electric field amplitudes $E_2=0.2-1$ for the shear rates of $\dot{\gamma}=0.0001$ and $\dot{\gamma}=0.001$.}
%     \label{fig:dielec}
% \end{figure}
% \begin{figure}
%     \centering
%     \includegraphics[width=\linewidth]{Pictures/tauf.pdf}
%     \caption{The relaxation time of the charge sequence calculated using \eqref{eq:Polar2} at the electric field amplitudes $E_2=0.2-1$ for the shear rates of $\dot{\gamma}=0.0001$ and $\dot{\gamma}=0.001$.}
%     \label{fig:tauf}
% \end{figure}
% \begin{figure}
%     \centering
%     \includegraphics[width=\linewidth]{Pictures/tau.pdf}
%     \caption{The overall relaxation time calculated using \eqref{eq:Polar3} at the electric field amplitudes $E_2=0-1$ for the shear rates of $\dot{\gamma}=0.0001$ and $\dot{\gamma}=0.001$.}
%     \label{fig:tau}
% \end{figure}

\begin{figure}
    \centering
    \sidesubfloat[]{\includegraphics[width=0.8\linewidth]{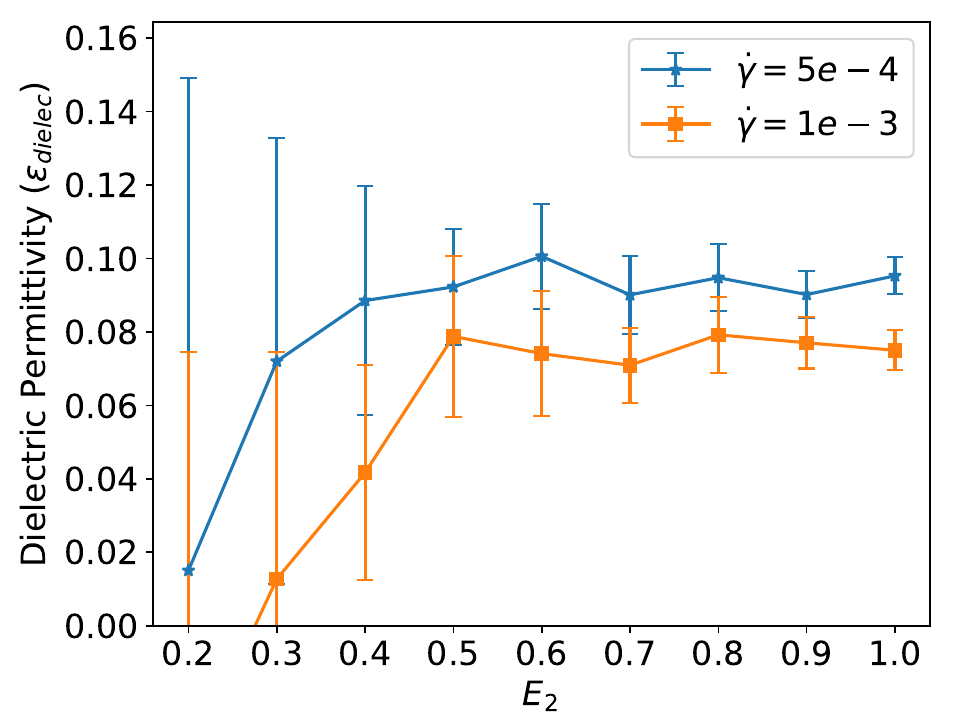}}\\
    \sidesubfloat[]{\includegraphics[width=0.8\linewidth]{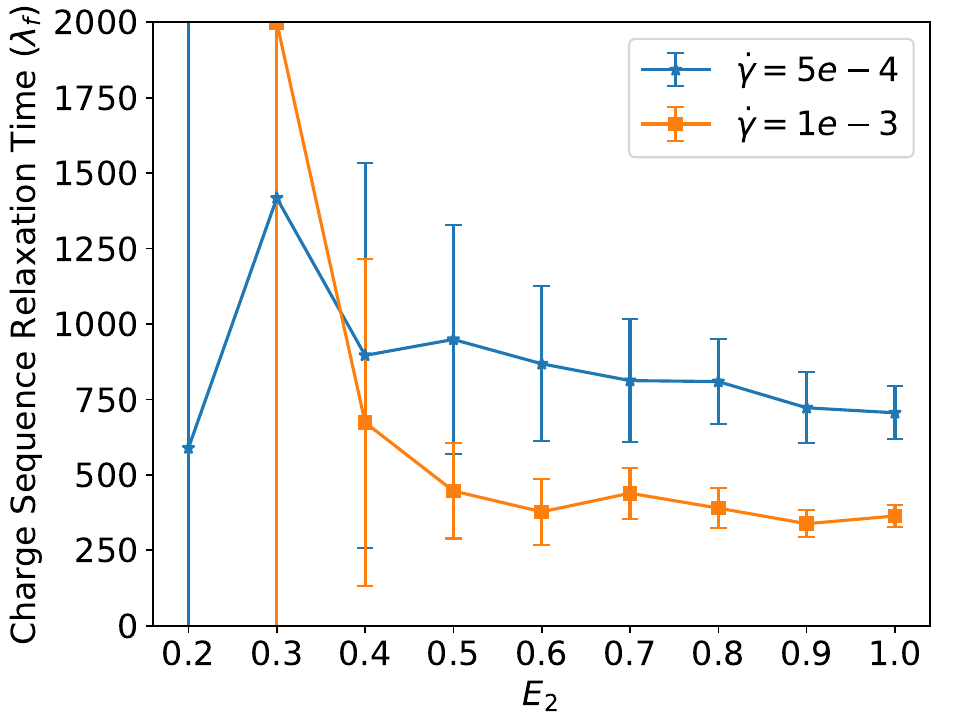}}\\
    \sidesubfloat[]{\includegraphics[width=0.8\linewidth]{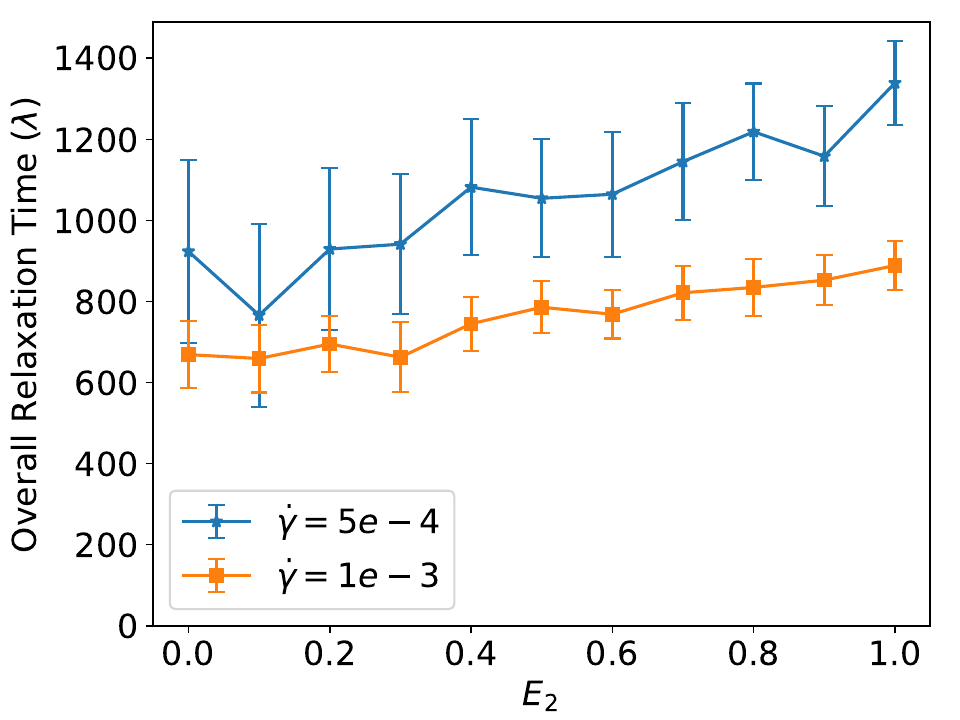}}
    \caption{(a) The dielectric permittivity calculated using \eqref{eq:Polar1}, (b) the relaxation time of the charge sequence calculated using \eqref{eq:Polar2}, and (c) the overall relaxation time calculated using \eqref{eq:Polar3} at the electric field amplitudes $E_2=0-1$ for the shear rates of $\dot{\gamma}=5e-4$ and $\dot{\gamma}=1e-3$. All material functions are in LJ units. Error bars represent one standard deviation from the mean.}
    \label{fig:Non-Linear-Mat-Func}
\end{figure}

Here we compare with the results from the coarse-grained MD simulations at a constant shear rate, once a steady-state response is reached, as shown in Fig. \ref{fig:CouettePolarFits}.
Using a non-linear, least squares best fits from the data (assuming the general form of the equation $y=A E_2^2+B$ where A and B are fitted coefficients), the material properties based off the polarizations stresses shown in (\ref{eq:Polar1})-(\ref{eq:Polar2}) at a shear rate of $\dot{\gamma}=5e-4$ are $\lambda_f = 700$ and $\epsilon_{dielec}=0.0964$. Similarly at a shear rate of $\dot{\gamma}=1e-3$, we estimate $\lambda_f = 341$ and $\epsilon_{dielec}=0.0826$.
These material functions present an overall decrease relative to the shear rate.
When comparing the cases of $\dot{\gamma}=5e-4$ and $\dot{\gamma}=1e-3$, the Weissenberg numbers are $Wi=1.93$ and $Wi=3.86$ using the longest relaxation from the Rouse chain geometry, generally meaning that non-linear effects relative to shear rate are expected.
Thus, we interpret the material functions as state dependent and not merely constants.
   
The polarization of $\sigma_{11}-\sigma_{33}$ gives an insight into how $\sigma_{11}$ should present the quadratic scaling of $E_2$ that is dependent on $\dot{\gamma}^2$, which becomes prevalent in $\dot{\gamma}= 1e-3$ as shown in Fig. \ref{fig:CouettePolarFits}b.
%Fig. \ref{fig:CouettePolarFits}b also demonstrates the limitation of the model, where the second normal stress difference is found to be non-zero in the case of $E_2=0$.
%The original UCM model does not predict the second normal stress difference, therefore low shear rates should be applied to the UCEM model when assessing its applicability.
Fig. \ref{fig:CouettePolarFits}b also reveals a limitation: the MD data show a non-zero second normal stress difference at $E_2 = 0$, a nonlinear effect absent from the UCM model. 
This limits the applicability of the UCEM to shear rates low enough that such higher-order contributions remain negligible.

One can further find the state-dependent behavior of the material functions of each constant shear rate for the dielectric permittivity and the relaxation times, as shown in Fig. \ref{fig:Non-Linear-Mat-Func}. 
Interestingly, we observe a non-linear dependence of the effective dielectric permittivity and charge sequence relaxation timescale as a function of the electric field, which plateaus at moderate to large field strengths. 
This initial non-linearity of both shear cases comes from the negative second normal stress difference shown within Fig. \ref{fig:CouettePolarFits}, but plateaus off once the electric field force is the largest contributing factor.
To emphasize, these negative/initial values are not the actual permittivity and charge sequence relaxation time, but a consequence of the limitations of the MD model where charge-charge interactions become comparable to the charge-field interactions, as evident by the increasingly large error bars below $E < 0.4$. We estimate that charge-field interactions dominate when $E\gg Cq/r^2$, where $r$ is the characteristic distance between two charged beads. For the MD simulations studied here, $r\approx1\sigma$, and therefore charge-charge interactions are on the order of $U_C(r)\sim0.08\epsilon$. This explains the anomalous behavior of $\epsilon_{dielec}$ and $\lambda_f$ at the lower end of the field strengths studied.

When relating the dielectric permittivity to the charge sequence relaxation time, we find that for lower shear rates $\lambda_f$ is larger, but tends to stay constant at large electric field amplitudes. 
The charge sequence relaxation time remains relatively constant at high electric field intensities because the model assumes charges are fixed to specific monomers and interact with the electric field through a purely force-based mechanism.
Thus, the coupling term $\lambda_f\epsilon_{dielec}\stackrel{\nabla}{(E_iE_j)}$ can be reasonably approximated at low electric fields and shear rates to be zero, and the only electrostatic stress contribution needed is the Maxwell-stress tensor.
This suggests that the UCEM model becomes prominent at large electric fields and low shear rates.
Finally, $\lambda$ shows a clear increase in the overall relaxation time as the electric field increases for both cases, suggesting that the electric field also slows the relaxation of the polymer due to the charge sequence interactions (see Fig. \ref{fig:Non-Linear-Mat-Func}c).

\begin{figure}
    \centering
    \sidesubfloat[]{\includegraphics[width=0.8\linewidth]{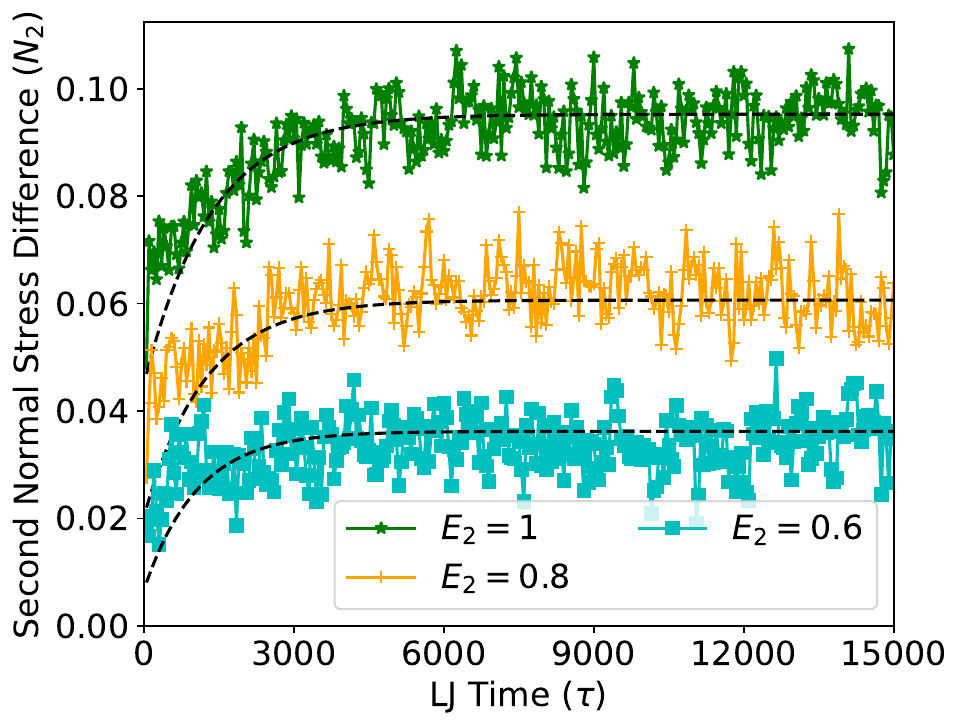}}\\
    \sidesubfloat[]{\includegraphics[width=0.8\linewidth]{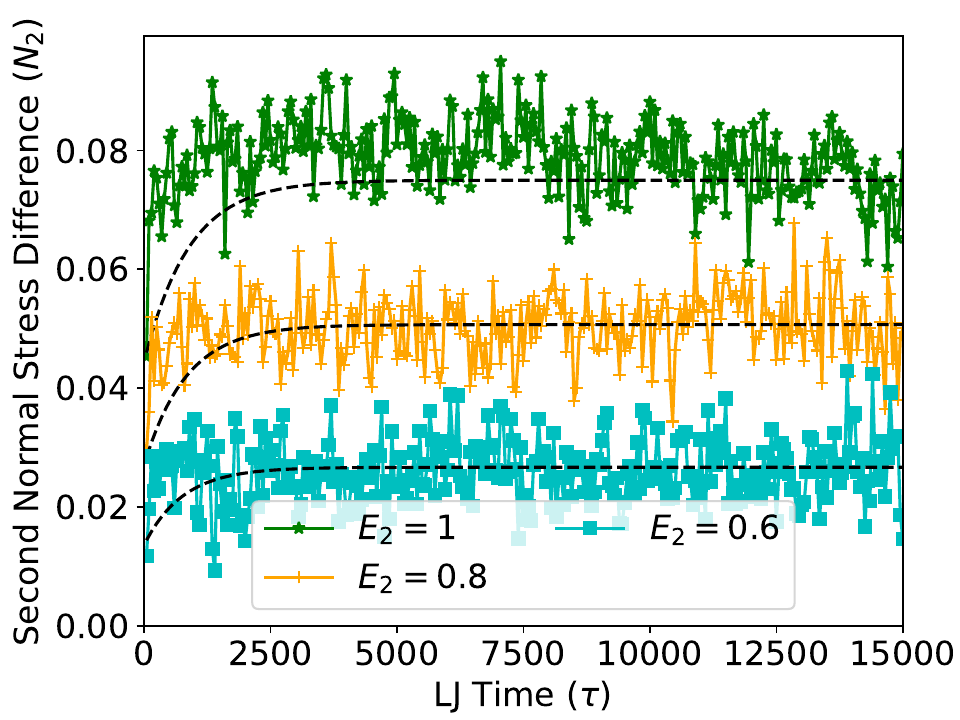}}
    \caption{The transient response of the second normal stress difference for (a) $\dot{\gamma}=5e-4$ and (b) $\dot{\gamma}=1e-3$ at $E_2=0.6,0.8,1$. The dashed black curves in both figures are the analytical predictions given by \eqref{eq:N2_Transient_E} using the material function values found in Fig. \ref{fig:Non-Linear-Mat-Func}.}
    \label{fig:N_2}
\end{figure}

\begin{figure}
    \centering
    \sidesubfloat[]{\includegraphics[width=0.8\linewidth]{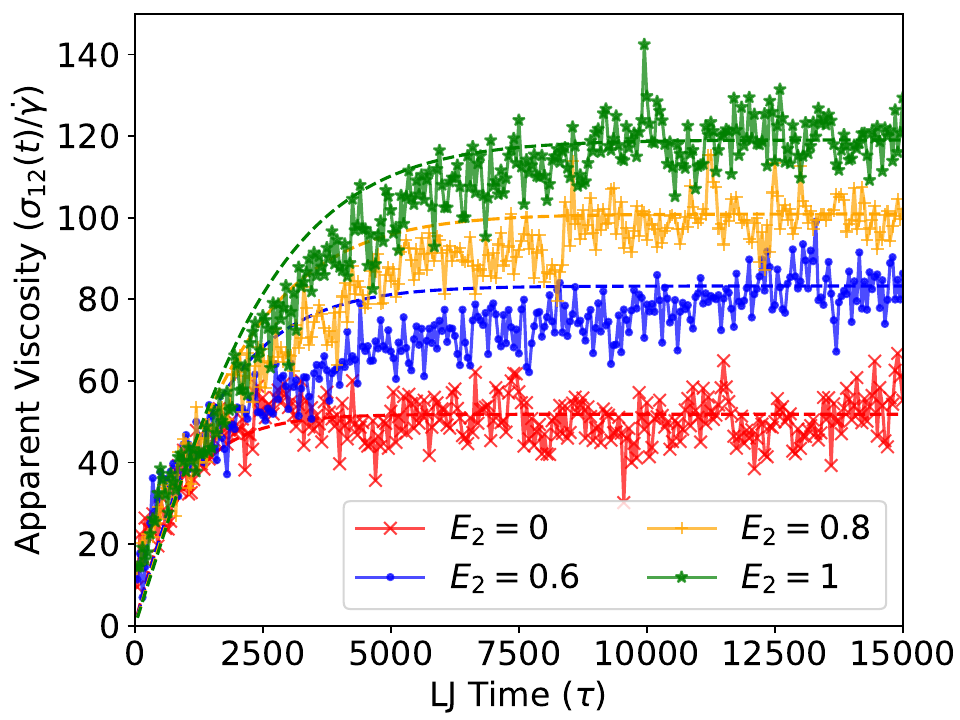}}\\
    \sidesubfloat[]{\includegraphics[width=0.8\linewidth]{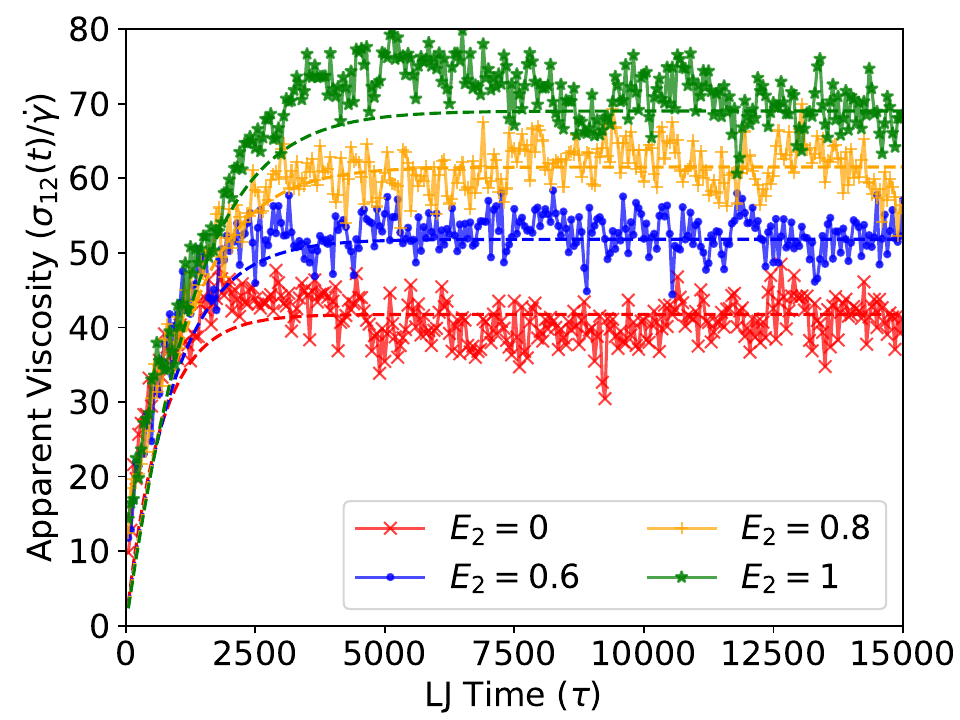}}
    \caption{Transient apparent viscosity response of a step strain rate of (a) $\dot{\gamma}=5e-4$, and (b) $\dot{\gamma}=1e-3$ at $E_2 = 0-1$. The dashed lines in both figures are the analytical predictions found in \eqref{eq:TransientFit} using the material function values found in Fig. \ref{fig:Non-Linear-Mat-Func}.}
    \label{fig:Transient}
\end{figure}

To validate the observed increase in the global relaxation time, we analyze the system's transient behavior to determine if the time required to reach a steady state extends as the electric field intensity increases.
However, for the transient behavior to be examined, we note that at $t=0$, a step function for both the shear rate $\dot{\gamma}(t)=\dot{\gamma}H(t)$ and electric field $E_2(t)=E_0H(t)$ are applied, such that the constitutive equations for the polymeric stress become,
\begin{equation}
    \begin{gathered}
        \sigma_{22}^p+\lambda\dot{\sigma}_{22}^p=-\lambda_f\epsilon_{dielec}E_0^2\delta(t)\\
        \sigma_{12}^p+\lambda\dot{\sigma}_{12}^p=(\lambda_f\epsilon_{dielec}E_0^2+\eta)\dot{\gamma}H(t)+\lambda\sigma_{22}^p\dot{\gamma}H(t)
    \end{gathered}
\end{equation}
To solve this, we first find $\sigma_{22}^p$, which was previously found in \eqref{eq:Transient_s_22_p} under a quasi-static medium.
With this solution, we can show the shear stress ordinary differential equation (ODE) as
\begin{equation}
    \sigma_{12}^p+\lambda\dot{\sigma}_{12}^p=\Big((\lambda_f\epsilon_{dielec}E_0^2+\eta)\dot{\gamma}-\lambda_f\epsilon_{dielec}E_0^2e^{-t/\lambda}\dot{\gamma}\Bigg)H(t),
\end{equation}
which has the solution
\begin{equation}
    \begin{gathered}
        \sigma_{12}^p(t)=\sigma_{12}(t) \\=(\eta+\lambda_f\epsilon_{dielec}E_0^2)\dot{\gamma}(1-e^{-t/\lambda})-\frac{\lambda_f}{\lambda}\epsilon_{dielec}E_0^2\dot{\gamma}te^{-t/\lambda}.\label{eq:TransientFit}
    \end{gathered}
\end{equation}
The viscoelastic response of the electric stress polarization in \eqref{eq:N2_Transient_E} can be determined in the second normal stress difference as shown in Fig. \ref{fig:N_2} for both shear rates tested, and also providing the overall relaxation time of the system from Fig. \ref{fig:Non-Linear-Mat-Func}c for $\dot{\gamma}=1e-3$.
Figure \ref{fig:Transient} presents the analytical curves derived from \eqref{eq:TransientFit} alongside the raw data, incorporating the global relaxation times determined from the $\sigma_{11}-\sigma_{33}$ response at $\dot{\gamma}=1e-3$ and $\dot{\gamma}=5e-4$. 
These results are further parameterized by the dielectric permittivity from Fig. \ref{fig:Non-Linear-Mat-Func}a and the optimized charge sequence relaxation time.
As illustrated in Fig. \ref{fig:Transient}, the data show strong agreement with the model and accurately capture the extended time required to reach a steady state. In contrast, the standard linear viscoelastic response, $\eta = \eta_0(1 - e^{-t/\lambda})$, fails to account for this delay and significantly underpredicts the transient duration.
This additional lag time to steady-state is due to the medium relaxing to its fully polarized state, which is required to capture the expected increase in the apparent viscosity.

% \begin{figure}
%     \centering
%     \includegraphics[width=\linewidth]{Pictures/N2_Long_0.001.pdf}
%     \caption{The transient response of the second normal stress difference for $\dot{\gamma}=0.001$ at $E_2=0.6,0.8,1$. The dashed black curve is the analytical prediction found in \eqref{eq:N2_Transient_E} using the best fit values $\tau_f=371$, and $\epsilon_{dielec}=0.0856$ with the separate overall relaxation times $\tau(E_2=0.6)=645$, $\tau(E_2=0.8)=673$, and $\tau(E_2=1)=728$.}
%     \label{fig:N2_Long_E1_0.001}
% \end{figure}
% \begin{figure}
%     \centering
%     \includegraphics[width=\linewidth]{Pictures/N2_long_0.0001.pdf}
%     \caption{The transient response of the second normal stress difference for $\dot{\gamma}=0.0001$ at $E_2=0.6,0.8,1$. The dashed black curve is the analytical prediction found in \eqref{eq:N2_Transient_E} using the steady-state dielectric permittivity found in Fig. \ref{fig:Non-Linear-Mat-Func}a and averaging the initial jump between time $t=50-250$ for $\tau_f/\tau$. The overall relaxation time is found using the value of $\tau_f/\tau$ with the best fit value of $\tau_f=1138$, where the overall relaxation times are found to be $\tau(E_2=0.6)=1803$, $\tau(E_2=0.8)=2028$, and $\tau(E_2=1)=2279$.}
%     \label{fig:N2_Long_E1_0.0001}
% \end{figure}

% \begin{figure}
%     \centering
%     \includegraphics[width=\linewidth]{Pictures/Transient_Long_0.001.pdf}
%     \caption{Transient shear stress response of a step strain rate of $\dot{\gamma}=0.001$ under  $E_2=0,0.5,0.8,1$ electric fields with a linear viscoelastic fit in \eqref{eq:TransientFit} using $\eta=29.75$, $\epsilon_{dielec}=0.0856$, and $\tau_f=371$. The overall relaxation time used for each electric field case from the normal stresses was $\tau(E_2=0)=521$, $\tau(E_2=0.5)=633$, $\tau(E_2=0.8)=673$, and $\tau(E_2=1)=728$.}
%     \label{fig:Transient_0.001}
% \end{figure}

% \begin{figure}
%     \centering
%     \includegraphics[width=\linewidth]{Pictures/Transient_Long_0.0001.pdf}
%     \caption{Transient shear stress response of a step strain rate of $\dot{\gamma}=0.0001$ under  $E_2=0,0.6,0.8,1$ electric fields with a linear viscoelastic fit in \eqref{eq:TransientFit} using $\eta=40.49$, $\tau_f=1135$, and $\epsilon_{dielec}$ were taken from Fig. \ref{fig:Non-Linear-Mat-Func}a. The overall relaxation time used for each electric field case were taken from Fig. \ref{fig:N2_Long_E1_0.0001}.}
%     \label{fig:Transient_0.0001}
% \end{figure}

\subsection{Uniaxial extensional flow}

Extensional flow, particularly uniaxial extensional flow, serves as a standard benchmark for evaluating the constitutive response of polymers and is fundamental to high-value manufacturing applications such as electrospinning.
The kinematic definitions that ensure incompressibility for this specific flow behavior are given by,
\begin{equation}
    v_i=
    \begin{bmatrix}
       \dot{\epsilon}x_1\\-\dot{\epsilon}x_2/2 \\-\dot{\epsilon}x_3/2  
    \end{bmatrix},
\end{equation}
\begin{equation}
    v_{i,j}=\begin{bmatrix}
        \dot{\epsilon}&0 &0 \\
        0& -\dot{\epsilon}/2& 0\\
        0& 0&-\dot{\epsilon}/2
    \end{bmatrix},
\end{equation}
where $\dot{\epsilon}$ is the extensional rate.
By applying these kinematic definitions to the polymer constitutive response while holding the generality of all possible homogeneous, steady-state electric fields, the polymer stress comes out as 
\begin{equation}
    \begin{gathered}
        \sigma_{11}^{p}=\frac{2\dot{\epsilon}(\eta+\lambda_f\epsilon_{dielec}E_1^2)}{1-2\lambda\dot{\epsilon}},\\
        \sigma_{22}^{p}=-\frac{(\eta+\lambda_f\epsilon_{dielec}E_2^2)\dot{\epsilon}}{(1+\lambda\dot{\epsilon})} ,\\
        \sigma_{33}^{p}=-\frac{(\eta+\lambda_f\epsilon_{dielec}E_3^2)\dot{\epsilon}}{(1+\lambda\dot{\epsilon})} ,\\
        \sigma_{12}^{p}=\frac{\lambda_f\epsilon_{dielec}E_1E_2}{2-\lambda\dot{\epsilon}}\dot{\epsilon} ,\\
        \sigma_{13}^{p}=\frac{\lambda_f\epsilon_{dielec}E_1E_3}{2-\lambda\dot{\epsilon}}\dot{\epsilon} ,\\
        \sigma_{23}^{p}=-\frac{\lambda_f\epsilon_{dielec}E_2E_3}{1+\lambda\dot{\epsilon}}\dot{\epsilon} ,\\
    \end{gathered}
\end{equation}
where the total stress is then the sum of the electrostatic stress contributions and the pressure tensor
\begin{equation}
    \begin{gathered}
        \sigma_{11}=\frac{2\dot{\epsilon}(\eta+\lambda_f\epsilon_{dielec}E_1^2)}{1-2\lambda\dot{\epsilon}}+\frac{\epsilon_{dielec}}{2}(E_1^2-E_2^2-E_3^2)-p,\\
        \sigma_{22}=-\frac{(2\eta+\lambda_f\epsilon_{dielec}E_2^2)\dot{\epsilon}}{2(1+\lambda\dot{\epsilon})} +\frac{\epsilon_{dielec}}{2}(E_2^2-E_1^2-E_3^2)-p,\\
        \sigma_{33}=-\frac{(2\eta+\lambda_f\epsilon_{dielec}E_3^2)\dot{\epsilon}}{2(1+\lambda\dot{\epsilon})} +\frac{\epsilon_{dielec}}{2}(E_3^2-E_2^2-E_1^2)-p,\\
        \sigma_{12}=\frac{\lambda_f\epsilon_{dielec}E_1E_2}{2-\lambda\dot{\epsilon}}\dot{\epsilon} +\epsilon_{dielec}E_1E_2,\\
        \sigma_{13}=\frac{\lambda_f\epsilon_{dielec}E_1E_3}{2-\lambda\dot{\epsilon}}\dot{\epsilon} +\epsilon_{dielec}E_1E_3,\\
        \sigma_{23}=-\frac{\lambda_f\epsilon_{dielec}E_2E_3}{1+\lambda\dot{\epsilon}}\dot{\epsilon} +\epsilon_{dielec}E_2E_3.\\
    \end{gathered}
\end{equation}
In the original UCM model, a singularity appears for extensions at $\dot{\varepsilon}=\frac{1}{2\lambda}$ that is non-physical in nature, but is a mathematical limitation of the model under high extensional flows \cite{macosko1994rheology}.
When examining the UCEM model response, a similar singularity forms within the total stress $\sigma_{11}$ from the polymer contribution, along with another singularity in both $\sigma_{12}$ and $\sigma_{13}$ at $\dot{\epsilon}=\frac{2}{\lambda}$ due to certain electric field contributions.
However, this new singularity arises at a larger extensional rate than $\sigma_{11}$, indicating that the model will break down before this becomes apparent.
Furthermore, the $\lambda_f$ term acts as a mechanism for anisotropic viscosity enhancement across all stress components; unlike its $\lambda$ counterpart, it does not induce singularity formations within the constitutive response.

\begin{figure}
    \centering
    \sidesubfloat[]{\includegraphics[width=0.8\linewidth]{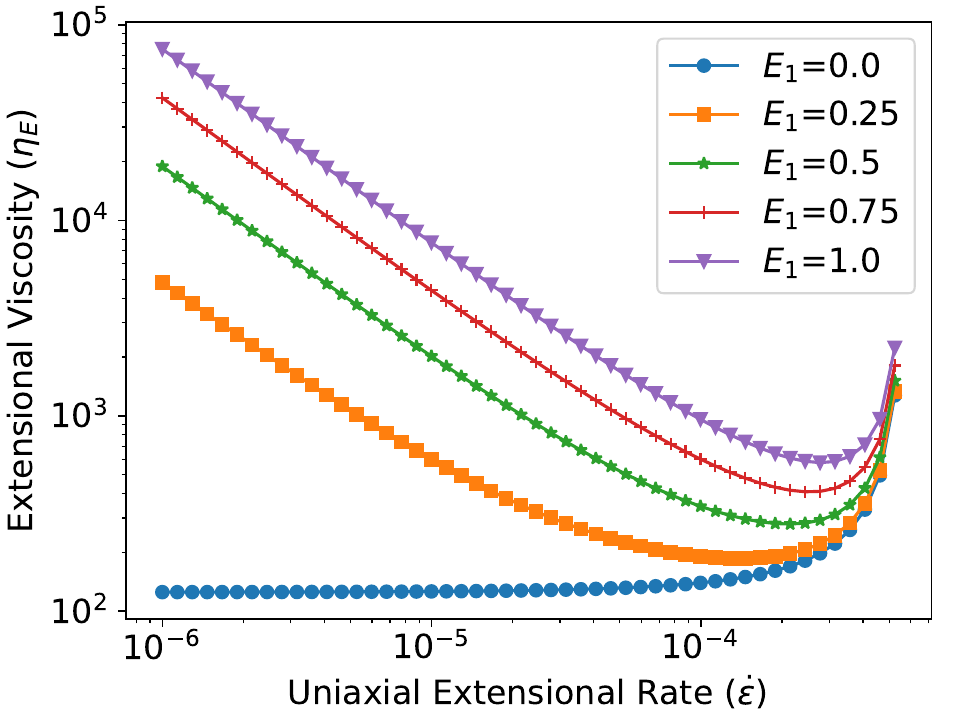}}\\
    \sidesubfloat[]{\includegraphics[width=0.8\linewidth]{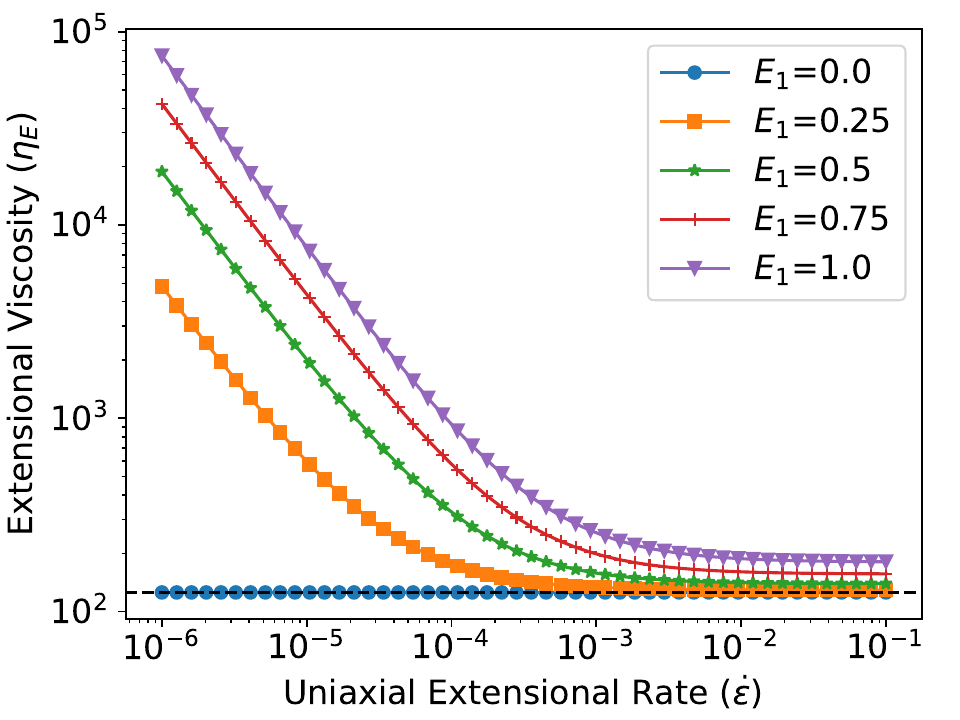}}
    \caption{(a) Extensional viscosity $\eta_E$ vs. extensional rate $\dot{\epsilon}$ at electric fields ranging from $E_1=0-1$ with the material functions $\eta = 41.75$, $\epsilon_{dielec} = 0.075$, $\lambda=888.76$, and $\lambda_f=363.7$. (b) Extensional viscosity $\eta_E$ vs. extensional rate $\dot{\epsilon}$ at electric fields ranging from $E_1=0-1$ for a $\lambda\approx0$ with the material functions $\eta = 41.75$, $\epsilon_{dielec} = 0.075$, and $\lambda_f=363.7$. The black dashed line represents the lower limit of the extensional viscosity of $\eta_E=3\eta$.}
    \label{fig:Linear_Extension}
\end{figure}

Typically, during electro-spinning, only $E_1$ is non-zero, giving the extensional viscosity
\begin{equation}
    \eta_E = \frac{\sigma_{11}-\sigma_{22}}{\dot{\epsilon}}=\frac{2(\eta+\lambda_f\epsilon_{dielec}E_1^2)}{1-2\lambda\dot{\epsilon}}+\frac{\eta}{(1+\lambda\dot{\epsilon})}+\frac{\epsilon_{dielec}E_1^2}{\dot{\epsilon}}.
\end{equation}
In Fig. \ref{fig:Linear_Extension}, we plot $\eta_E$ as a function of $\dot{\epsilon}$ for two values of $\lambda$ under applied fields using material functions $\eta$, $\epsilon_{dielec}$ and $\lambda_f$ measured from Fig. 4 at $E=1$. At low extensional rates far from the singularity, the electric field is found to enhance the extensional response through two distinct mechanisms: a yield-like behavior scaling as $\frac{\epsilon_{dielec}E_1^2}{\dot{\epsilon}}$ and a viscous-type contribution governed by $\lambda_f$. Furthermore, in the case that $\lambda$ is small (where finite rotations are small relative to the flow), but with a large $\epsilon_{dielec}E_iE_j$ (finite rotations are large relative to the electric field), the extensional viscosity becomes stable in the form
\begin{equation}
    \eta_E = 3\eta+2\lambda_f\epsilon_{dielec}E_1^2+\frac{\epsilon_{dielec}E_1^2}{\dot{\epsilon}},
\end{equation}
where the extensional viscosity at high extensional rates plateaus at $\eta_E=3\eta+2\lambda_f\epsilon_{dielec}E_1^2$ as shown in Fig. \ref{fig:Linear_Extension}.

\subsection{Steady-state pressure-driven flow}

For a two-dimensional flow with a singular constant pressure differential under Stokes flow conditions, the general force balance comes out to be
\begin{equation}
    \sigma_{11,1}+\sigma_{12,2}=p_{,1},
\end{equation}
\begin{equation}
    \sigma_{12,1}+\sigma_{22,2}=0.
\end{equation}
For a pressure-driven flow, the only dependence of the total stresses is $\sigma_{ij}\equiv \sigma_{ij}(x_2)$
\begin{equation}
    \sigma_{12,2}=p_{,1},
\end{equation}
\begin{equation}
    \sigma_{22,2}=0.
\end{equation}
One can then solve for the upper-convected derivatives for a steady state flow with the velocity dependence $(v_1(x_2),v_2=0)$
\begin{equation}
    \stackrel{\nabla}{\sigma_{ij}^p}
    =-v_{1,2}\begin{bmatrix}
        2\sigma_{12}^p & \sigma_{22}^p \\ \sigma_{22}^p & 0
    \end{bmatrix},
\end{equation}
\begin{equation}
   \stackrel{\nabla}{(E_iE_j)}
    =-v_{1,2}\begin{bmatrix}
        2E_1E_2 & E_2E_2 \\ E_2E_2 & 0
    \end{bmatrix}.
\end{equation}
With this, the polymer stress contributions in matrix form are found to be
\begin{equation}
\begin{gathered}
    \begin{bmatrix}
        \sigma_{11}^p &\sigma_{12}^p\\\sigma_{12}^p&\sigma_{22}^p
    \end{bmatrix}
    -\lambda v_{1,2}\begin{bmatrix}
        2\sigma_{12}^p & \sigma_{22}^p \\ \sigma_{22}^p & 0
    \end{bmatrix}
    \\=\eta\begin{bmatrix}
        0 &v_{1,2}\\v_{1,2}&0
    \end{bmatrix}
    +\lambda_f \epsilon_{dielec}v_{1,2}\begin{bmatrix}
        2E_1E_2 & E_2E_2 \\ E_2E_2 & 0
    \end{bmatrix}
    \end{gathered}
\end{equation}
which can be written out individually as
\begin{equation}
    \begin{gathered}
        \sigma_{11}^p=2\lambda\sigma_{12}v_{1,2}+2\lambda_fv_{1,2}E_1E_2,\\
        \sigma_{22}^p= 0,\\
        \sigma_{12}^p=(\eta+\lambda_f\epsilon_{dielec}E_2E_2)v_{1,2}.
    \end{gathered}
\end{equation}

\begin{figure}
    \centering
    \includegraphics[width=0.8\linewidth]{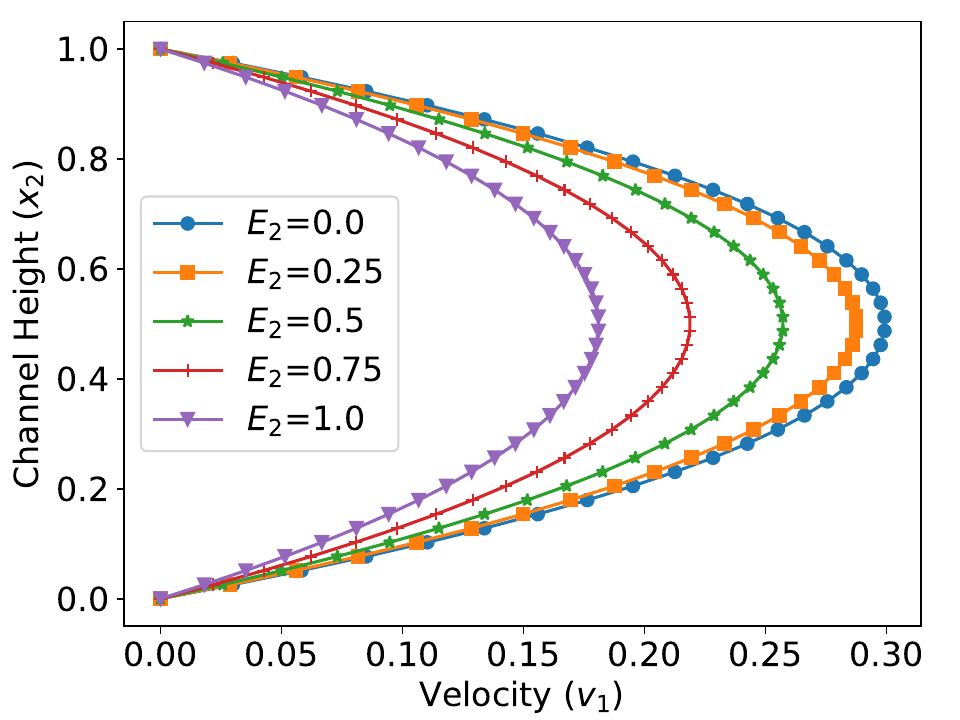}
    \caption{Velocity profiles $v_1(x_2)$ of pressure-driven flows at electric fields ranging from $E_2=0-1$ with the material functions $\eta = 41.75$, $\epsilon_{dielec} = 0.075$, $\lambda=888.76$, and $\lambda_f=363.7$ and a pressure differential of $p_{,1}=-100$.}
    \label{fig:pressure}
\end{figure}

Using the restrictions placed by the force balance (the electric fields are assumed constant, meaning that the electrostatic contribution to the total stress decomposition is zero), the general solution for the velocity profile is found to be
\begin{equation}
    \sigma_{12,2}=(\eta+\lambda_f\epsilon_{dielec}E_2E_2)v_{1,22}=p_{,1},
\end{equation}
\begin{equation}
    v_{1,22}=\frac{p_{,1}}{\epsilon_{dielec}\lambda_fE_2E_2+\eta},
\end{equation}
\begin{equation}
    v_1=\frac{p_{,1}}{\lambda_f\epsilon_{dielec}E_2E_2+\eta}\frac{x_2^2}{2}+Ax_2+B.
\end{equation}
Applying no-slip boundary conditions,
\begin{equation}
    \begin{gathered}
    v_1(x_2=0)=0=\frac{p_{,1}}{\lambda_f\epsilon_{dielec}E_2E_2+\eta}\frac{0^2}{2}+A(0)+B\\
    \implies B=0,
    \end{gathered}
\end{equation}
\begin{equation}
    \begin{gathered}
    v_1(x_2=h)=0=\frac{p_{,1}}{\lambda_f\epsilon_{dielec}E_2E_2+\eta}\frac{h^2}{2}+Ah\\
    \implies A=-\frac{p_{,1}}{\lambda_f\epsilon_{dielec}E_2E_2+\eta}\frac{h}{2}.
    \end{gathered}
\end{equation}
We note that polyelectrolyte chains can exhibit slip at boundaries under combined flow and electric fields \cite{LI20191066,WANG2023124545}; however, here we adopt standard no-slip conditions as a baseline, independent of the charge distribution. The velocity profile with no-slip boundaries is thus
\begin{equation}
    \begin{gathered}
    v_1=\frac{p_{,1}}{\lambda_f\epsilon_{dielec}E_2E_2+\eta}\frac{x_2^2}{2}-\frac{p_{,1}}{\lambda_f\epsilon_{dielec}E_2E_2+\eta}\frac{x_2h}{2}\\
    =\frac{p_{,1}}{\lambda_f\epsilon_{dielec}E_2E_2+\eta}\Bigg(\frac{x_2^2-x_2h}{2} \Bigg).
    \end{gathered}
\end{equation}
A visual representation of this flow profile can be seen in Fig. \ref{fig:pressure}.
Consistent with the observations in Couette flow, the coupling between the electric field and the fluid results in a thickening response that manifests as an increased apparent viscosity, leading to a reduced flow rate under a finite $E_2$ field. The resulting velocity profile is comparable to the experimental findings,\cite{abu-jdayil_effects_1996} where polar silica gel dispersions in paraffin oil similarly exhibited a quadratic velocity profile that thickened with increasing $E_2$ intensity.

\subsection{Small amplitude oscillatory shear flow}
For a small amplitude oscillatory shear (SAOS), the shear of the fluid is assumed in the form \cite{macosko1994rheology}
\begin{equation}
    \gamma=\gamma_0 \sin\left(\omega t\right), 
\end{equation}
where $\gamma_0$ is the amplitude of the oscillatory shear, and $\omega$ is the frequency of the oscillation.
The shear rate is then found as
\begin{equation}
    \dot{\gamma}=v_{1,2}=\gamma_0\omega \cos\left(\omega t\right).
\end{equation}
One can now define the shear stress and shear stress rate as
\begin{equation}
    \begin{gathered}
        \sigma_{12}=\lambda_0'\sin(\omega t)+\lambda_0^{''}\cos(\omega t),\\
        \dot{\sigma}_{12}=\omega\lambda_0'\cos(\omega t)-\omega\lambda_0^{''}\sin(\omega t),
    \end{gathered}
\end{equation}
where $\lambda_0'=G'\gamma_0$ and $\lambda_0^{''}=G^{''}\gamma_0$ are the storage and loss modulus, respectively, pre-normalization with the shear amplitude.
Applying this form to \eqref{eq:ElectroMaxwellModel} while assuming only the shear stress is time dependent, one finds the form of the balance equation as
\begin{equation}
    \begin{gathered}
    (\lambda_0'-\omega \lambda_0^{''}\lambda)\sin(\omega t)+(\lambda_0'\lambda\omega+\lambda_0^{''})\cos(\omega t)\\=(\lambda_f\epsilon_{dielec}E_2E_2+\eta)\gamma_0\omega \cos(\omega t)+ \epsilon_{dielec}E_1E_2.
    \end{gathered}
\end{equation}
When solving for the coefficients in front of the $\sin$ and $\cos$ terms, the following relations appear
\begin{equation}
    \begin{gathered}
        G'=\omega\lambda G^{''},\\
        G^{''}+\omega\lambda G'=(\lambda_f\epsilon_{dielec}E_2E_2+\eta)\omega,
    \end{gathered}
\end{equation}
where the storage and loss moduli can be defined as
\begin{equation}
    \begin{gathered}
        G'=\frac{\omega^2\lambda}{(1+\omega^2\lambda^2)}(\lambda_f\epsilon_{dielec}E_2E_2+\eta),\\
        G^{''}=\frac{\omega}{(1+\omega^2\lambda^2)}(\lambda_f\epsilon_{dielec}E_2E_2+\eta).
    \end{gathered}
\end{equation}
Finally, the shift in phase between the strain and stress is defined as
\begin{equation}
    \tan(\delta)=\frac{G''}{G^{'}}=\frac{1}{\omega\lambda},
\end{equation}
meaning that a constant electric field does not affect the phase angle shift for the linear regime.
These findings align with recent experimental results, which demonstrated that PMMA melts exhibit no phase shift under constant $E_2$ fields during SAOS experiments within the linear regime across various temperatures.\cite{huo_electric_2020} It should be noted that the current continuum model is not designed to predict non-linear shear responses and, therefore, is not intended to characterize the full spectrum of shear amplitudes observed experimentally.

\section{Conclusion}
\paragraph*{Summary.}
In this work, we model the electro-viscoelastic effect of charged polymers using analytical and numerical methods spanning multiple scales. A modified viscoelastic Maxwell fluid model -- the \emph{upper-convected electro-Maxwell (UCEM) model} -- is developed, incorporating a polarization stress coupled to a charge sequence relaxation time within the polymer chain. The UCEM model predictions for Couette flow are validated against (1) stresses derived from a modified Rouse model with distributed charges, and (2) viscoelastic polarization responses from coarse-grained molecular dynamics simulations of Kremer-Grest chains with charges along the backbone. Dynamic predictions are also consistent with experimental observations, with the electric field producing no change in the phase shift between shear stress and strain.
\paragraph*{Limitations.}
The UCEM model rests on two principal simplifications. First, the underlying Rouse-level description neglects long-range electrostatic charge-charge interactions. This restricts the model's direct applicability to salt-free melts of polyampholytes with net-neutral charge sequences, where the dominant multipole is dipolar and interactions are effectively short-ranged ($\sim 1/r^3$), in contrast to the long-range ($\sim 1/r$) monopole interactions that govern polyelectrolyte behavior. Electrorheological phenomena driven by strong, unscreened charge-charge coupling therefore lie outside the present scope. Second, the model is linear in shear rate and therefore cannot capture the shear thinning commonly exhibited by polymer solutions and melts at higher shear rates. These simplifications are deliberate: the UCEM model represents the simplest continuum formulation that couples Rouse relaxation modes to the frequencies of charge distributions and characterizes the competing effects of flow and electric fields on the polymer chain conformations.
\paragraph*{Outlook.}
With steady-state validation established, a natural extension is to incorporate transient electric fields and examine the competing effects of polarization and relaxation timescales. Of particular interest is the response under alternating electric fields, where the interaction between the AC frequency and the charge sequence relaxation time may reveal additional physics. More broadly, extensions to nonlinear viscoelastic constitutive forms and incorporation of charge-charge interactions represent important directions for expanding the model's applicability. Lastly, this model offers insights for electrically assisted manufacturing processes such as electrospinning and electrospraying. By demonstrating how field-induced anisotropic viscosity enhancement can be manipulated, the framework suggests new methods for controlling material properties such as fiber diameter and coating uniformity. Ultimately, linking molecular charge patterns directly to macroscopic flow behavior establishes a design tool for creating sequence-defined polymers with optimized processing characteristics.

\begin{acknowledgments}
The authors thank Joseph Peterson (University of California, Los Angeles) for helpful discussions. This research was supported in part by an appointment to the Department of Defense (DoD) Research Participation Program
administered by the Oak Ridge Institute for Science and Education (ORISE) through an interagency agreement between the U.S.
Department of Energy (DOE) and the DoD. 
ORISE is managed by ORAU under DOE contract number DE-SC0014664. All opinions
expressed in this paper are the authors' and do not necessarily reflect the policies and views of DoD, DOE, or ORAU/ORISE.
JE and MG acknowledge the support of the Air Force Research Laboratory.
\end{acknowledgments}

\section*{Author Declarations}
\subsection*{Conflict of Interest}
The authors have no conflicts to disclose.

\subsection*{Author Contributions}
\textbf{Zachary Wolfgram}: Data curation (lead); Conceptualization (lead); Formal Analysis (lead); Software (lead); Writing/Review and Editing (Primary/equal).
\textbf{Jeffrey Ethier}: Project administration (equal); Supervision (equal); Writing/Review and Editing (secondary/equal).
\textbf{Matthew Grasinger}:  Project administration (equal); Supervision (equal);  Writing/Review and Editing (secondary/equal).

\section*{Data Availability Statement}
The data that support the findings of this study are available from the corresponding author upon reasonable request.

\appendix

\section{Derivation of Electro-Viscoelastic Stress from the Modified Rouse Model}
\label{app:RouseDoi}

For simplicity, and for demonstrating objectivity and anisotropy in the electric field interactions, one can use the normalized Langevin equations shown in \eqref{eq:Lag_Equations} with dual shear rates and electric field potential contributions shown as,
\begin{equation}
    \dot{X}_{px}=-\frac{X_{px}}{\tau_p}+\frac{q_pE_x}{\zeta_p}+\dot{\gamma}_1X_{py}+g_{px}
\end{equation}
\begin{equation}
    \dot{X}_{py}=-\frac{X_{py}}{\tau_p}+\frac{q_pE_y}{\zeta_p}+\dot{\gamma}_2X_{px}+g_{py}
\end{equation}
\begin{equation}
    \dot{X}_{pz}=-\frac{X_{pz}}{\tau_p}+\frac{q_pE_z}{\zeta_p}+g_{pz}
\end{equation}
where $X_{px}$, $X_{py}$, and $X_{pz}$ are the normalized positions of the chain in the $x$, $y$, and $z$ plane, respectively, $\dot{\gamma}_1=\partial v_x/\partial y$ and $\dot{\gamma}_2=\partial v_y/\partial x$ are the shear rates along the $x$ and $y$ direction, and $q_pE_x$ and $q_pE_z$ represents the directional forces under a constant electric field with any mode $p$ cosine charge sequence.  

For determining the shear stress the following terms are required,
\begin{equation}
    \begin{gathered}
    \langle{X_{px}}\rangle=\int^t_{-\infty}e^{\frac{-(t-t')}{\tau_p}}\left(\frac{q_{p}E_x}{\zeta_p}+\dot{\gamma}_1X_{py}+g_{px}\right)dt'\\\approx \frac{\tau_p q_pE_x}{\zeta_p}+\tau_p\dot{\gamma}_1\frac{q_{p}E_y}{\zeta_p} + O(\dot{\gamma}^2),
    \end{gathered}
\end{equation}
\begin{equation}
    \begin{gathered}
    \langle{X_{py}}\rangle=\int^t_{-\infty}e^{\frac{-(t-t')}{\tau_p}}\left(\frac{q_{p}E_y}{\zeta_p}+\dot{\gamma}_2X_{px}+g_{py}\right)dt'\\\approx \frac{\tau_p q_{p}E_y}{\zeta_p}+\tau_p\dot{\gamma}_2\frac{q_{p}E_x}{\zeta_p} + O(\dot{\gamma}^2).
    \end{gathered}
\end{equation}
where the brackets $\langle{...}\rangle$ denotes the ensemble average. One can retrieve the final terms needed for the shear stress in (\ref{eq:avg_stress_Rouse}),
\begin{equation}
    \begin{gathered}
    \langle{X_{py}X_{py}}\rangle=\int^{t'}_{-\infty}\int^{t}_{-\infty}e^{\frac{-(t+t'-t_1-t_2)}{\tau_p}}\Bigg(\frac{q_pE_y}{\zeta_p}+\dot{\gamma}_2X_{px}+g_{py}\Bigg)\\\Bigg(\frac{q_pE_y}{\zeta_p}+\dot{\gamma}_2X_{px}+g_{py}\Bigg)dt_1dt_2\\
    \approx \bigg(\frac{\tau_pq_{p}E_y}{\zeta_p}\bigg)^2 +\frac{k_BT}{k_p} + 2\frac{q_{p}^2\tau_p^2E_xE_y}{\zeta^2_p}\dot{\gamma_2} + O(\dot{\gamma}^2)
    \end{gathered}
\end{equation}

\begin{equation}
    \begin{gathered}
    \langle{X_{px}X_{px}}\rangle=\int^{t'}_{-\infty}\int^{t}_{-\infty}e^{\frac{-(t+t'-t_1-t_2)}{\tau_p}}\Bigg(\frac{q_{p}E_x}{\zeta_p}+\dot{\gamma}_1X_{py}+g_{px}\Bigg)\\
    \Bigg(\frac{q_{p}E_x}{\zeta_p}+\dot{\gamma}_1X_{py}+g_{px}\Bigg)dt_1dt_2
    \\
    \approx \bigg(\frac{\tau_pq_{p}E_x}{\zeta_p}\bigg)^2 +\frac{k_BT}{k_p} + 2\frac{q_{p}^2\tau_p^2E_xE_z}{\zeta^2_p}\dot{\gamma_1} + O(\dot{\gamma}^2).
    \end{gathered}
\end{equation}

To obtain $\langle X_{px}X_{py}\rangle$, the following can be done with the initial Langevin equations,\cite{doi1996introduction}
\begin{equation}
    \dot{X}_{px}X_{py}=-\frac{X_{px}X_{py}}{\tau_p}+\frac{q_{p}E_xX_{py}}{\zeta_p}+\dot{\gamma}_1X_{py}X_{py}+g_{px}X_{py},
\end{equation}
\begin{equation}
    \dot{X}_{py}X_{px}=-\frac{X_{py}X_{px}}{\tau_p}+\frac{q_{p}E_yX_{px}}{\zeta_p}+\dot{\gamma}_2X_{px}X_{px}+g_{py}X_{px},
\end{equation}
%which, added together, shows
%\begin{equation}
%    \begin{gathered}
%    \dot{u}_pX_{py}+\dot{w}_pu_p=\dot{X_{py}u_p}\\
%    =-\frac{2u_pX_{py}}{\tau_p}+\frac{F_{px}^EX_{py}}{\zeta_p}+\dot{\gamma}_1X_{py}X_{py}+g_{px}X_{py}\\+\frac{F_{pz}^Eu_p}{\zeta_p}+\dot{\gamma}_2u_pu_p+g_{pz}u_p.
%    \end{gathered}
%\end{equation}
then adding and integrating with respect to time, the average comes out to be
\begin{equation}
    \begin{gathered}
        \langle X_{px}X_{py}\rangle=\frac{\tau_p}{2}\Bigg(2\frac{\tau_pq_{p}^2E_xE_y}{\zeta_p^2}+\frac{k_BT}{k_p}(\dot{\gamma}_2+\dot{\gamma}_1)\\+2\frac{\tau_p^2}{\zeta_p^2}({q_{p}E_x}^2\dot{\gamma}_2+{q_{p}E_y}^2\dot{\gamma}_1) + O(\dot{\gamma}^2)\Bigg).
    \end{gathered}
\end{equation}

\section{Derivation of Electro-Viscoelastic Stress from the Modified Elastic Dumbbell}
\label{app:Dumbbell} 
To further motivate the upper-convected derivative of the electric field dyadic in the UCEM model, we derive the stress evolution for an elastic dumbbell with equal and opposite charges on each end, following the framework of Larson\cite{larson1999structure}. This minimal microstructural model produces the same $\stackrel{\nabla}{(E_i E_j)}$ coupling that appears in the continuum formulation, demonstrating that the convected derivative arises naturally from the kinematics of a polarized elastic element in flow.
For an elastic dumbbell under a constant electric field with equal and opposite charges on each end of the dumbbell, the force balance comes out to
\begin{equation}
    \frac{D({R}_i)}{Dt} = \kappa_{ik}R_k - \frac{4k_BT\beta^2}{\zeta}R_i-\frac{2k_BT}{\zeta}\frac{\partial \Xi}{\partial R_i} + \frac{2q}{\zeta}E_i,
\end{equation}
where $R_i$ is the end-to-end vector, $\kappa_{ij}=v_{i,j}$, $k_BT$ is the thermal energy, $\zeta$ is the friction coefficient, $\beta^2 = \frac{3}{2N_kb_k^2}$ with $N_k$ being the number of links on a chain of length $b_K$, $\Xi$ is the distribution of conformations, and $2q$ is the magnitude difference of the charges on each end, such that the charge end to end measure is $q_2-q_1=q-(-q)=2q$.
If one were to include the Coulomb force between the two charges, the additional term in the force balance would be
\begin{equation}
    F^{coul}_i = \frac{2q^2}{4\pi\epsilon\zeta R_kR_k} \frac{R_i}{\sqrt{R_pR_p}},
\end{equation}
where $\epsilon$ is the permittivity of the medium and $\frac{R_i}{\sqrt{R_pR_p}}$ is the normal vector of the end-to-end vector.
Now, this term will always be present within the medium, whether that be under a flow or an electric field, and will only produce an isotropic, non-zero equilibrium length depending on the strength of the charges and the spring between the charges.
For our analysis, we want to examine non-zero orientations that arise due to the charge alignment under an external electric field and along with the flow field, and thus remove the contribution to simplify the system.
Using the Smoluchowski equation that describes the change of the probability distribution $(\Xi)$ as a function of the end-to-end vector
\begin{equation}
    \frac{D\Xi}{Dt} + \frac{\partial}{\partial R_m}\Bigg(\frac{D{R}_m}{Dt} \Xi \Bigg) = 0,
\end{equation}
the force balance can be introduced to show
\begin{equation}
    \dot{\Xi} + \frac{\partial}{\partial R_i}\Bigg(\Bigg(\kappa_{ik}R_k - \frac{4k_BT\beta^2}{\zeta}R_i-\frac{2k_BT}{\zeta}\frac{\partial \Xi}{\partial R_i} + \frac{2q}{\zeta}E_i\Bigg) \Xi\Bigg) = 0.
\end{equation}
We introduce two evolution equation forms
\begin{equation}
    \begin{gathered}
        \frac{D\Xi}{Dt}R_j + \frac{\partial}{\partial R_i}\Bigg(\Bigg(\kappa_{ik}R_k - \frac{4k_BT\beta^2}{\zeta}R_i \\-\frac{2k_BT}{\zeta}\frac{\partial \Xi}{\partial R_i} + \frac{2q}{\zeta}E_i\Bigg) \Xi\Bigg)R_j = 0,
    \end{gathered}
\end{equation}
\begin{equation}
    \begin{gathered}
        \frac{D\Xi}{Dt}R_jR_p + \frac{\partial}{\partial R_i}\Bigg(\Bigg(\kappa_{ik}R_k \\
        - \frac{4k_BT\beta^2}{\zeta}R_i-\frac{2k_BT}{\zeta}\frac{\partial \Xi}{\partial R_i} + \frac{2q}{\zeta}E_i\Bigg) \Xi\Bigg)R_jR_p = 0
    \end{gathered}
\end{equation}
where we are interested in finding the average end-to-end vector $\langle R_j\rangle=\int R_j\Xi dR^3$, and the average conformation tensor $\langle R_jR_p\rangle=\int R_jR_p\Xi dR^3$.
Applying integration by parts on the 2nd term in both evolution equations, the final forms of each simplify to
\begin{equation}
    \begin{gathered}
        \frac{D\langle R_jR_p\rangle}{Dt} - \kappa_{jk}\langle R_kR_p\rangle - \langle R_jR_k\rangle\kappa_{pk}+\frac{8k_BT\beta^2}{\zeta}\langle R_jR_p\rangle \\- \frac{2q}{\zeta}(\langle R_j\rangle E_p+E_j\langle R_p\rangle) = \frac{4k_BT}{\zeta}\delta_{jp},
    \end{gathered}
\end{equation}
\begin{equation}
    \frac{D \langle R_j\rangle}{Dt} = \kappa_{jk}\langle R_k\rangle -\frac{4k_BT\beta^2}{\zeta}\langle R_j\rangle + \frac{2q}{\zeta}E_j.
\end{equation}
Using the form of $\langle R_j\rangle$ and substituting into the evolution of $\langle R_jR_p\rangle$, and using the definition of the upper-convected derivative $\Big(\stackrel{\nabla}{\langle R_jR_p\rangle}=\frac{D\langle R_jR_p\rangle}{Dt} - \kappa_{jk}\langle R_kR_p\rangle - \langle R_jR_k\rangle\kappa_{pk}\Big)$, one can show
\begin{equation}
    \begin{gathered}
        \stackrel{\nabla}{\langle R_jR_p\rangle} + \frac{8k_BT\beta^2}{\zeta}\langle R_jR_p\rangle+\frac{q}{2k_BT\beta^2}\Bigg(\frac{D(\langle R_j\rangle E_p + E_j \langle R_p\rangle)}{Dt} \\-\kappa_{jk}\langle R_k\rangle E_p - E_j\langle R_k\rangle \kappa_{pk} \Bigg) = \frac{4k_BT}{\zeta}\delta_{pj} + \frac{2q^2}{\zeta k_BT\beta^2} E_pE_j. \label{eq:Appendix_R_iR_j}
    \end{gathered}
\end{equation}
To remove the average end-to-end vector from the evolution, one can argue that $\langle R_j\rangle$ is a steady state quantity, such that its original evolution equation becomes
\begin{equation}
    \langle R_j\rangle = \frac{\zeta}{4k_BT\beta^2}\Bigg(\kappa_{jk}\langle R_k\rangle +\frac{2q}{\zeta}E_j\Bigg), \label{eq:Appen_B_R}
\end{equation}
where the electric field induces a non-zero average of the end-to-end vector, which the flow kinematics can further influence.
An example of this is for a $v_{1,2}$ Couette flow, where any electric field arrangement gives the steady-state vector
\begin{equation}
    [\langle R_j\rangle] = \frac{2q}{4k_BT\beta^2}\begin{bmatrix}
        \frac{\zeta}{4k_BT\beta^2}v_{1,2}E_2 + E_1 \\
        E_2 \\
        E_3
    \end{bmatrix}.
\end{equation}

Using \eqref{eq:Appen_B_R} and enforcing a linear response with the shear rate where $\dot{\gamma}^2 \approx 0$, the configuration tensor evolution equation simplifies to
\begin{equation}
    \begin{gathered}
        \stackrel{\nabla}{\langle R_jR_p\rangle} + \frac{8k_BT\beta^2}{\zeta}\langle R_jR_p\rangle+\frac{q^2}{4k_B^2T^2\beta^4}\Bigg(-\kappa_{jk}E_kE_p - E_jE_k\kappa_{pk} \Bigg) \\= \frac{4k_BT}{\zeta}\delta_{pj} + \frac{2q^2}{\zeta k_BT\beta^2} E_pE_j. 
    \end{gathered}
\end{equation}
Finally, the electric field and flow coupling $-\kappa_{kj}E_kE_p - E_jE_k\kappa_{kp}$ is not an objective measure, but can be derived from the upper-convected derivative, such that the conformation evolution of the dumbbell model under a constant electric field becomes
\begin{equation}
    \begin{gathered}
        \stackrel{\nabla}{\langle R_jR_p\rangle} + \frac{8k_BT\beta^2}{\zeta}\langle R_jR_p\rangle+\frac{q^2}{4k_B^2T^2\beta^4}\stackrel{\nabla}{(E_jE_p)} \\= \frac{4k_BT}{\zeta}\delta_{pj} + \frac{2q^2}{\zeta k_BT\beta^2} E_jE_p, 
    \end{gathered}
\end{equation}
where $\frac{D(E_jE_p)}{Dt}=0$ for steady-state and homogeneous electric fields.
This same arrangement can also be found if one approximates the average end-to-end vector as $\langle R_i\rangle\approx\frac{2q}{4k_BT\beta^2}E_i$ and substituted into \eqref{eq:Appendix_R_iR_j}.
Converting from the average configuration tensor to the polymer stress with $\sigma_{ij}^p=2vk_BT\beta^2\langle R_iR_j \rangle$ where v is the density of dumbbells, finds the form
\begin{equation}
    \begin{gathered}
        \stackrel{\nabla}{\sigma^p_{jp}} + \frac{1}{\lambda}\sigma_{jp}^p+\epsilon_{dielec}\stackrel{\nabla}{(E_jE_p)} = \frac{G}{\lambda}\delta_{jp} + \frac{\epsilon_{dielec}}{\lambda} E_jE_p, 
    \end{gathered}
\end{equation}
where $G=vk_BT$, $\lambda = \frac{\zeta}{8k_BT\beta^2}$, and $\epsilon_{dielec}=\frac{vq^2}{2k_BT\beta^2}$.

\section*{references}
\bibliographystyle{unsrt}
\doublespacing
\bibliography{sample}% Produces the bibliography via BibTeX.

\end{document}